\documentclass[twocolumn,prb,superscriptaddress,longbibliography]{revtex4-2}

\usepackage{amsmath,amsfonts,amssymb}
\usepackage{graphicx,color}
\usepackage{hyperref}
\usepackage{float}
\usepackage{multirow}
\usepackage{hhline}
\usepackage{ulem}
\usepackage{mathtools}
\usepackage{siunitx}
\usepackage[caption=false]{subfig}

\newcommand{\ket}[1]{\left|#1\right\rangle}

\makeatletter
\renewcommand{\thetable}{\arabic{table}}
\renewcommand{\fnum@table}{Table \thetable}

\@ifundefined{textcolor}{}
{%
 \definecolor{BLACK}{gray}{0}
 \definecolor{WHITE}{gray}{1}
 \definecolor{RED}{rgb}{1,0,0}
 \definecolor{GREEN}{rgb}{0,1,0}
 \definecolor{BLUE}{rgb}{0,0,1}
 \definecolor{CYAN}{cmyk}{1,0,0,0}
 \definecolor{MAGENTA}{cmyk}{0,1,0,0}
 \definecolor{YELLOW}{cmyk}{0,0,1,0}
}

\makeatother

\begin{document}

\title{A Cascaded Random Access Quantum Memory}
\author{Ziqian Li}
\altaffiliation{These authors contributed equally to this work}
\affiliation{Department of Physics and Applied Physics, Stanford University, Stanford, California 94305, USA}
\affiliation{Department of Physics, University of Chicago, Chicago, Illinois 60637, USA}
\affiliation{SLAC National Accelerator Laboratory, Menlo Park, California 94025, USA}

\author{Eesh Gupta}
\altaffiliation{These authors contributed equally to this work}
\affiliation{Department of Physics and Applied Physics, Stanford University, Stanford, California 94305, USA}
\affiliation{SLAC National Accelerator Laboratory, Menlo Park, California 94025, USA}

\author{Fang Zhao}
\affiliation{Fermi National Accelerator Laboratory, Batavia, IL 60510, USA}

\author{Riju Banerjee}
\affiliation{Department of Physics, University of Chicago, Chicago, Illinois 60637, USA}

\author{Yao Lu}
\affiliation{Fermi National Accelerator Laboratory, Batavia, IL 60510, USA}

\author{Tanay Roy}
\affiliation{Fermi National Accelerator Laboratory, Batavia, IL 60510, USA}

\author{Andrew Oriani}
\affiliation{Department of Physics, University of Chicago, Chicago, Illinois 60637, USA}

\author{Andrei Vrajitoarea}
\affiliation{Center for Quantum Information Physics, Department of Physics, New York University, New York 10003, USA}

\author{Srivatsan Chakram}
\affiliation{Department of Physics and Astronomy, Rutgers University, Piscataway, New Jersey 08854, USA}

\author{David I. Schuster}
\affiliation{Department of Physics and Applied Physics, Stanford University, Stanford, California 94305, USA}
\affiliation{SLAC National Accelerator Laboratory, Menlo Park, California 94025, USA}

\date{\today}

\begin{abstract}
 Dynamic random access memory (DRAM) is critical to classical computing but notably absent in current superconducting quantum processors. Integrating high-coherence memory units would enable resource-efficient control of logical qubits and allow the separate optimization of logic and storage subsystems. Here, we realize an 8-bit cascaded random access quantum memory (RAQM). By introducing a buffer layer between the processor and a multimode storage cavity, we leverage the control resources of a single transmon to address eight memory modes while isolating them from processor non-linearities. We demonstrate arbitrary random access with an average infidelity of $\lesssim 1.5\%$ per mode, characterizing the many-body interactions that dominate the error budget. This architecture enables a significant reduction in control lines per logical qubit and supports transversal operations within the memory module, establishing a scalable unit cell for fault-tolerant quantum architectures.

\end{abstract}

\maketitle

\section{Introduction}

As experimental demonstrations of quantum error correction (QEC) advance toward fault-tolerant operation~\cite{Ofek2016, Takeda2022, ryan_anderson_ftqec_2021, Sivak2023, Bluvstein2024, Acharya2024}, the field faces a critical challenge: scaling the hardware resources required to control increasingly large numbers of logical qubits.  Current architectures typically rely on a uniform grid of qubits where control lines scale linearly with qubit count.  In classical computing, this scaling bottleneck is resolved by architectural specialization, with a small active processor and a large random access memory (RAM), separating logic from information storage to maximize density and efficiency.  Similarly, realizing large-scale quantum processors will require dedicated, high-capacity quantum memories to complement logical processing units~\cite{dave_memory_propsal, Duckering2020, girvin_strategies_tradeoffs_memory_2024}.

Developing such a memory is particularly urgent for superconducting circuits.  Unlike atomic systems, which possess naturally high coherent hyperfine states that serve as intrinsic memories~\cite{willke_2018_hyperfine, Bluvstein2024}, superconducting qubits lack a natural storage subsystem.  The physical requirements for processing and storage are fundamentally conflicting: high-fidelity gates require strong nonlinearity and interaction (e.g. transmons ~\cite{koch2007transmon}, fluxonium~\cite{Manucharyan2009Fluxonium}), whereas long-lived storage favors linear, weakly interacting systems (e.g. superconducting cavities~\cite{romanenko_cavities, milul_cavity_qubit, oriani2024niobiumcoaxialcavitiesinternal}, nanomechanical resonators~\cite{MacCabe2020NanoAcoustic, Beccari2022, nanomech_ultra_high, Cupertino2024}, or spin ensembles~\cite{Ruskuc2022NuclearSpinWave}).  Alternatively, noise protected qubits~\cite{noise_protected_review_gyenis, bifluxon_Kalashnikov_2020, zero_pi} promise long coherence in their protected state but may have slower gates than simpler qubits. A memory architecture resolves these conflicting demands by integrating long-lived memories with non-linear processor qubits, enabling the separate optimization of logic and coherence.

Here, we realize a ``cascaded'' random access quantum memory (RAQM) module for superconducting circuits.  In this architecture, each processor qubit is integrated with a memory unit consisting of multiple storage modes.  This approach allows the control resources of a single logical qubit to be multiplexed across $N$ memory layers, significantly expanding logical capacity without increasing the control line count (Fig.~\ref{fig:figure1}a).  Crucially, this uses classical addressing to swap logical states between the storage layer and the processor and does not require coherent quantum addressing as in QRAM~\cite{qram_giovannetti, QRAM_3d_cav_weiss, router_scq_miao}.  Beyond hardware efficiency, the all-to-all connectivity within an RAQM module enables transversal gates between memory layers, reducing the time overhead for multi-logical-qubit operations compared to edge-to-edge connectivity constraints~\cite{Horsman_2012, zhou2025lowoverheadtransversalfaulttolerance, Duckering2020, dave_memory_propsal}.

We implement a single unit cell of  this architecture using a multimode 3D superconducting cavity system (Fig.~\ref{fig:figure1}b). We demonstrate universal control of this system, achieving a baseline swap infidelity of $< 0.5\%$ when accessing a single memory mode in isolation, a figure well below practical error-correction thresholds. Furthermore, we perform the first comprehensive benchmarking of the full RAQM module, characterizing the additional error overhead imposed by many-body decoherence between memory modes. We find that even in the presence of these interactions, the memory can be accessed in arbitrary order with an infidelity of $\lesssim 1.5\%$ per mode. Consequently, the aggregate error rate for memory sizes up to 7 modes remains below the depolarization threshold ($~17\%$) of the surface code~\cite{wootton_2012_high_threshold, bravyi_decoding_surface_code_2014, Criger2018multipathsummation, Kuo2022}, establishing the RAQM as a viable, scalable unit cell for fault-tolerant superconducting quantum computers.

\section{Assembling the Memories}
\begin{figure}[t]
    \centering
    \includegraphics{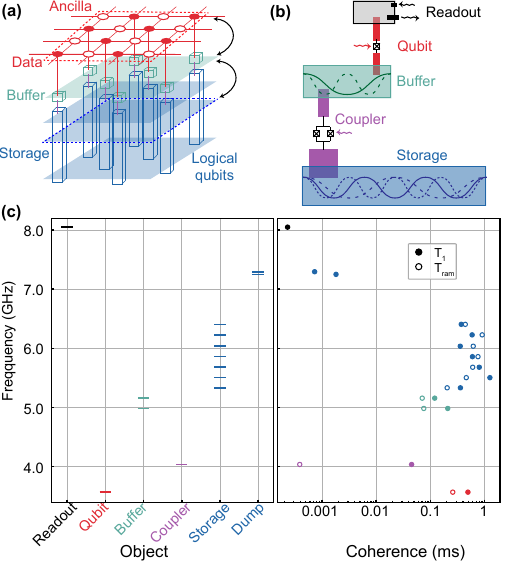}
    \caption{\textbf{(a)} Proposed fault-tolerant architecture using Random Access Quantum Memory (RAQM). Each data qubit in the transmon processor (red) also serves as a register qubit for its corresponding RAQM. Each RAQM consists of a buffer mode (green) and a storage multimode (blue). Modes from different cavities are organized into layers to store logical qubits. Logical qubits are transferred from storage layers (blue) to the buffer layer for processing. 
    \textbf{(b)} RAQM device schematic. The readout (black), buffer (green, $2$ modes), and storage (blue, $7$ modes) 3D flute cavities are fabricated on the same high-purity aluminum bulk. The transmon (red) is dispersively coupled to the readout and buffer cavities, enabling universal single-buffer-mode control. The RF-flux-modulated SQUID coupler (purple) enables parametrically driven beam-splitter interactions between coupled cavity modes, transferring information between buffer and storage layers. 
    \textbf{(c)} Frequency spectrum and mode coherence. The storage modes are evenly distributed between $5.3\,\mathrm{GHz}$ and $6.4\,\mathrm{GHz}$, with the highest $T_1$ exceeding $1.2\,\mathrm{ms}$ (see Supplementary Section~I). Error bars are smaller than the marker size. Details about dump modes are provided in Supplementary Section~II.}
    \label{fig:figure1}
\end{figure}

RAQM device comprises two layers (see Fig.~\ref{fig:figure1}(b)): a storage layer (S) and a buffer layer (B), with the buffer connected to a transmon processor for readout and gates. Both layers are implemented using 3D superconducting flute cavities, which are engineered to host multiple high-coherence bosonic modes within a single physical component~\cite{PhysRevLett.127.107701_vatsan_highQ}. Within the larger architecture of Fig.~\ref{fig:figure1}(a), this device would function as a single RAQM module: many long-lived modes in the storage layer are serviced by a common processor that is accessed through the buffer layer.

These layers have different functions inside the RAQM. The buffer layer serves as the processor-facing interface, engineered to strongly couple with the nonlinear processor while maintaining sufficient coherence. As such, it acts as a staging area where quantum states are temporarily stored for fast, programmable control. On the other hand, the storage layer is a bank of weakly coupled quantum memories optimized for coherence rather than gate operations. In the full RAQM architecture, these storage modes play the role of randomly addressable memory cells that can be selectively updated by the processor.

The nonlinear processor in our device is a transmon (Q) that is capacitively coupled to the buffer modes. Together, the transmon and a chosen buffer mode form a processing unit that enables universal single-mode control through optimal-control or sideband techniques~\cite{heeres_grape_2017, Eickbusch2022, SNAP_2015, huang2025fastsidebandcontrolweakly}. All state preparation, gates, and measurements in this work are implemented by driving the transmon and buffer, avoiding directly exposing the storage modes to non-linearity. Within the broader architecture, this transmon would belong to a large grid of qubits that periodically refreshes the memory through QEC.

To mediate communication between the buffer and storage layers, we introduce an RF-flux-modulated SQUID coupler~\cite{luSchoelkopf2023}. By parametrically driving this coupler, we implement programmable beam-splitter interactions between the buffer mode and selected storage modes, enabling coherent swaps between a chosen storage cell and the buffer. This multiplexed control strategy enables a single processor to access many storage modes while keeping the memory layer largely idle and protected.

The physics of the combined processor-RAQM system is described by the Hamiltonian $H = H_0 + H_{\text{proc}} + H_{\text{b.s.}}(t) + H_{\text{Kerr}}$. The fundamental harmonic resources are described by $H_0 = \omega_b a_b^\dagger a_b + \sum_{j=1}^7 \omega_{s_j} a_{s_j}^\dagger a_{s_j}$, where $a_b$ ($a_{s_j}$) and $\omega_b$ ($\omega_{s_j}$) denote the annihilation operator and frequency of the buffer (storage) mode $j$. 

Universal control of these modes is enabled by the nonlinear transmon processor, which is dispersively coupled to the buffer mode:
\begin{equation}
    H_{\text{proc}} = \omega_q a_q^\dagger a_q + \frac{\alpha}{2} a_q^\dagger a_q^\dagger a_q a_q + \chi_{qb} a_q^\dagger a_q a_b^\dagger a_b,
\end{equation}
where $\omega_q$, $\alpha$, and $\chi_{qb}$ are the transmon frequency, anharmonicity, and dispersive shift, respectively. To enable memory access, the SQUID coupler is modulated to generate time-dependent beam-splitter interactions:
\begin{equation}
    H_{\text{b.s.}}(t) = \sum_{j} \frac{g_j(t)}{2} \left( a_b^\dagger a_{s_j} + a_b a_{s_j}^\dagger \right).
\end{equation}
Here, $g_j(t)$ represents the tunable coupling rate. Finally, residual nonlinearities from the transmon and coupler introduce self- and cross-Kerr interactions $H_{\text{Kerr}} = \sum_{m,n} \chi_{mn} a_m^\dagger a_m a_n^\dagger a_n$, where indices run over the set $\{b, s_1, \dots, s_7\}$. Due to the cascaded architecture, the largest buffer-storage cross-Kerr is below $3.5\,\mathrm{kHz}$, while storage-storage terms remain at the hundred-Hz level.

In this work, we encode quantum information in the single-photon subspace $\{\ket{0}, \ket{1}\}$ of the cavity modes. The same hardware is also compatible with more complex bosonic encodings~\cite{marios_bosonic_QEC_2016} and dual-rail architectures~\cite{teoh_2023_dual_rail,Daul_rail_levine_2024}, in which the buffer--storage connectivity enables hardware-efficient error correction and logical operations.

\begin{figure*}[t]
    \centering
    \includegraphics{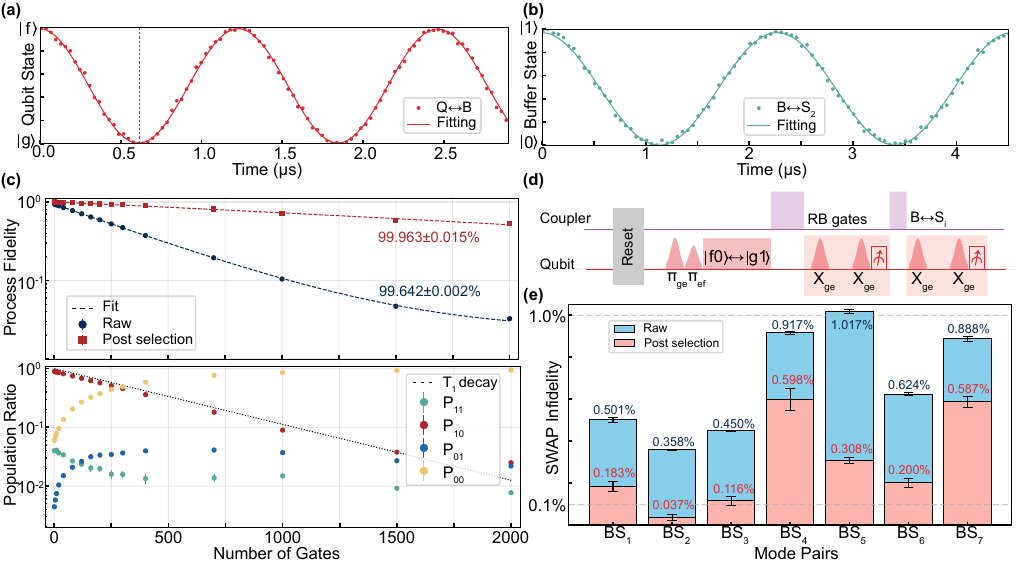}
    \caption{Single storage mode access. \textbf{(a)} Using four-wave mixing interactions $\ket{f0}\leftrightarrow\ket{g1}$ between the transmon and buffer mode, we prepare $\ket{1}$ in the buffer within $0.615\,\mu\mathrm{s}$. \textbf{(b)} By modulating the coupler's RF flux at the frequency difference between the buffer and storage mode $S_2$, a $0.439\,\mathrm{MHz}$ beam-splitter interaction is activated. \textbf{(c)} Randomized Benchmarking (RB) of the $B S_2$ swap. We prepare $\ket{1}$ in the buffer, execute RB sequences, and measure the joint buffer–storage state $\ket{B S_i}$. The top panel shows raw and post-selected probabilities of measuring $\ket{B S_2}$ returning to $\ket{10}$. Post-selection keeps outcomes with exactly one photon in the buffer–storage pair, isolating non-decoherence errors. The RB pulse sequences are shown in \textbf{(d)}. $\ket{B S_i}$ is extracted using two parity measurements and a swap, with the transmon actively reset to $\ket{g}$ between parity checks. Swap fidelities for all $B S_i$ are shown in \textbf{(e)}. RAQM reset protocols are discussed in Supplementary Section~III.}
    \label{fig:figure2}
\end{figure*}

\section{Controlling Single Quantum Memories}\label{sec: single mode memory access}
In this section, we will describe and benchmark each step of our RAQM's information transfer process. Accessing a specific memory mode proceeds in three steps: (1) a read operation swaps the quantum state from the target storage mode $S_i$ into the buffer $B$; (2) a gate operation modifies the state in the buffer via the transmon processor; and (3) a write operation swaps the modified state back into $S_i$. We characterize each stage of this cycle below.

We prepare states, manipulate them, and perform readout of the buffer mode through the transmon. Their dispersive interaction enables a four-wave mixing interaction $\ket{f0} \leftrightarrow \ket{g1}$ to realize the transmon-buffer swap~\cite{pechal_sideband_transmon_cavity}. Together with \(g-e\) transmon gates, we can perform universal gates on the buffer mode in its \(\{0,1\}\) subspace (See Supplementary Section VIII for details). As shown in Fig.~\ref{fig:figure2}(a), we prepare $\ket{1}$ in the buffer mode in $0.615\,\mu\mathrm{s}$ using this interaction. The transmon supports single-qubit gates with a fidelity of $99.96\%$ and single-shot readout with a fidelity of \(98.02\%\). To measure the buffer mode population, we use the transmon to perform the parity measurement, consisting of two transmon $\pi_{ge}/2$ rotations separated by a waiting time of $\pi/(\chi_{qb})$.

We realize buffer-storage swaps through the RF flux modulation of the SQUID coupler. To activate the beamsplitter interactions between any two modes, the external flux $\Phi_{\text{ext}}$ threading the SQUID loop is driven as $\Phi_{\text{ext}}=\Phi_{\text{DC}}+\epsilon\cos(\omega_dt+\phi)$, with DC bias \(\Phi_{DC} = 0.269\pi\) and modulation frequency $\omega_d$. By choosing the modulation phase ($\phi$) and amplitude ($\epsilon$) appropriately, we obtain the calibrated beamsplitter gate $U_{\text{b.s.}}(\phi)$. Fig.~\ref{fig:figure2}(b) shows the swap between the buffer mode and \(S_2\) storage mode with a rate of \(0.44\,\mathrm{MHz}\). A key innovation of this work is the on-chip, broadband fast-flux line designed for 3D-cavity systems. It preserves cavity mode coherence--achieving \(T_1\) up to \(1.25\,\mathrm{ms}\) (see Fig.~\ref{fig:figure1}(c))--while enabling fast beamsplitter interaction rates of \(0.3\text{--}0.5\,\mathrm{MHz}\) (see Supplementary Section~VI for details on the fast-flux-line design). In the present device, the swap rate is limited by the sideband frequency collisions between different storage-buffer pairs. Details about other buffer-storage swap rates are shown in the Supplementary Section II.

We quantify swap performance using beam-splitter randomized benchmarking (RB)~\cite{luSchoelkopf2023}. The sequence (Fig.~\ref{fig:figure2}(d)) initializes a single photon, applies random beam-splitter gates, and measures the joint parity. We define raw fidelity based on the recovery of the ideal state, and a post-selected fidelity by conditioning on the preservation of total photon number, $P_{10}/(P_{10}+P_{01})$, where $P_{nm}$ is the population of a buffer-storage pair having $n$ and $m$ photons. As shown in Fig.~\ref{fig:figure2}(e), we achieve an average raw (post-selected) fidelity of $99.32\%$ ($99.71\%$) across all $B-S_i$ pairs. The remaining incoherent error is attributed to coupler heating and cavity dephasing (see Supplementary Section X). While the current swap time ($\sim 0.5\,\mu\mathrm{s}$) is limited by frequency crowding, faster exchange rates are compatible with this architecture. Crucially, the measured swap infidelity satisfies even the stringent $\sim 1\%$ threshold typically required for active two-qubit gates in the surface code~\cite{Acharya2024}. Since memory access operations occur less frequently than active gates, this performance provides a substantial margin for fault-tolerant operation.

\section{Benchmarking Random Access Quantum Memory}\label{sec:raqm}

\begin{figure}[t]
    \centering
    \includegraphics{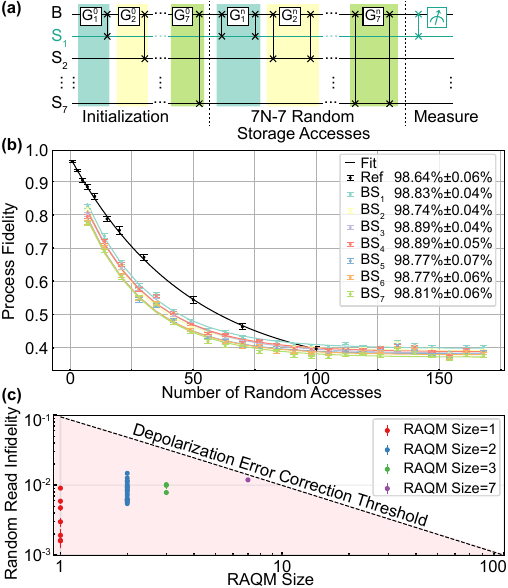}
    \caption{Random access quantum memory operation. \textbf{(a)} RAM RB circuit on $7$ storage modes. Each storage access gate is highlighted in a different color, containing a storage-buffer swap (read), a buffer gate $G_i^n$, and a buffer-storage swap (write). The storage access gate in the initialization stage does not contain the storage-buffer swap (read), as the storage modes are initialized in the ground state. At the end of the circuit, the target storage mode information is swapped into the buffer for readout. 
    \textbf{(b)} Measured random read fidelity for $7$ storage modes. The reference RB is the buffer mode's RB in the ${\ket{0},\ket{1}}$ subspace. \textbf{(c)} Measured average random read infidelity at different RAQM sizes. The region below the depolarization error correction~\cite{wootton_2012_high_threshold, bravyi_decoding_surface_code_2014, Criger2018multipathsummation, Kuo2022} threshold for surface code is highlighted.}
    \centering
    \label{fig:figure3}
\end{figure}

Having characterized the individual memory modes, we now evaluate the system’s performance as a random-access quantum memory (RAQM). An ideal RAQM enables high-fidelity read and write operations—implemented in our device as controlled buffer–storage swaps—on any selected mode while minimally disturbing all others. We benchmark these memory access operations across the full RAQM and discuss their implications for scalability.

To assess RAQM performance, we perform multiplexed RB across all memory modes, as shown in Fig.~\ref{fig:figure3}(a). The benchmarking sequence consists of the cyclic application of memory access gates, where full memory access consists of a buffer-storage swap (read), a buffer gate $G_j^n$, and a storage-buffer swap (write).  Each storage mode tracks its respective RB gate sequence during this multiplexing process. For smaller RAQM sizes, RB gates are applied cyclically to the corresponding subsystem.

For a size-7 RAQM, one memory cycle consists of 14 gates: two (read and write) applied on a selected storage mode $S_i$, and 12 applied to other storage modes. Ideally, the latter are effectively identity gates on \(S_i\).  However, due to decoherence and modes' cross-Kerrs, these identity gates have nonzero infidelities. From the RAQM RB data, we derive the random read fidelity $\bar{F}_{BS_i}$, which quantifies the average error per gate (Supplementary Section IX). As shown in Fig.~\ref{fig:figure3}(b), the measured fidelity $\bar{F}_{BS_i}$ for \(S_i\) ranges between $(98.74\%\pm0.04\%)$ and $(98.89\%\pm0.05\%)$. Overall, any read/write operation on a size-7 RAQM imparts an error \(\sim 1.2 \%\) on each mode.

These access errors necessitate periodic refresh of the memory using quantum error correction. We therefore benchmark the RAQM in QEC-relevant terms. Specifically, we identify the largest memory size for which errors accumulated over a full memory cycle remain below the QEC threshold. We address this by extracting the random read infidelity for different RAQM sizes (See Fig.~\ref{fig:figure3}(c)). In our device, a size-7 memory is the largest RAQM that remains below the surface-code depolarization threshold of about $17\%$~\cite{wootton_2012_high_threshold, bravyi_decoding_surface_code_2014, Criger2018multipathsummation, Kuo2022}. After a full cycle, which consists of a read and a write on each storage mode, the infidelity per mode is $1-0.9874^{14}=16.27\%$.

For purposes of benchmarking the memory, the threshold above assumes ideal processor gates and stabilizer measurements, with errors arising only from memory read/write operations. Notably, this threshold is higher than that for two-qubit gates, as memory swaps are performed only twice per QEC round. Larger RAQM requires more operations to complete one cycle, resulting in the power-law decay of the depolarization error threshold (See Fig.~\ref{fig:figure3}(c)).

These experiments reveal insights into the scalability of the RAQM architecture shown in Fig.~\ref {fig:figure1}(a). An RAQM substantially reduces the number of control lines only when the coherence of the memory exceeds that of the processor and the processor–memory swap error remains below the threshold. Increasing the number of memory modes enhances the logical qubit capacity without requiring additional control lines, but also amplifies idle errors. Beyond a certain memory size, the resources required to correct these additional errors can outweigh the control-line savings. Thus, suppressing idle errors is crucial for RAQM scalability. The optimal memory size is discussed further in Supplementary Section XII.

\begin{figure*}[t]
    \centering
    \includegraphics{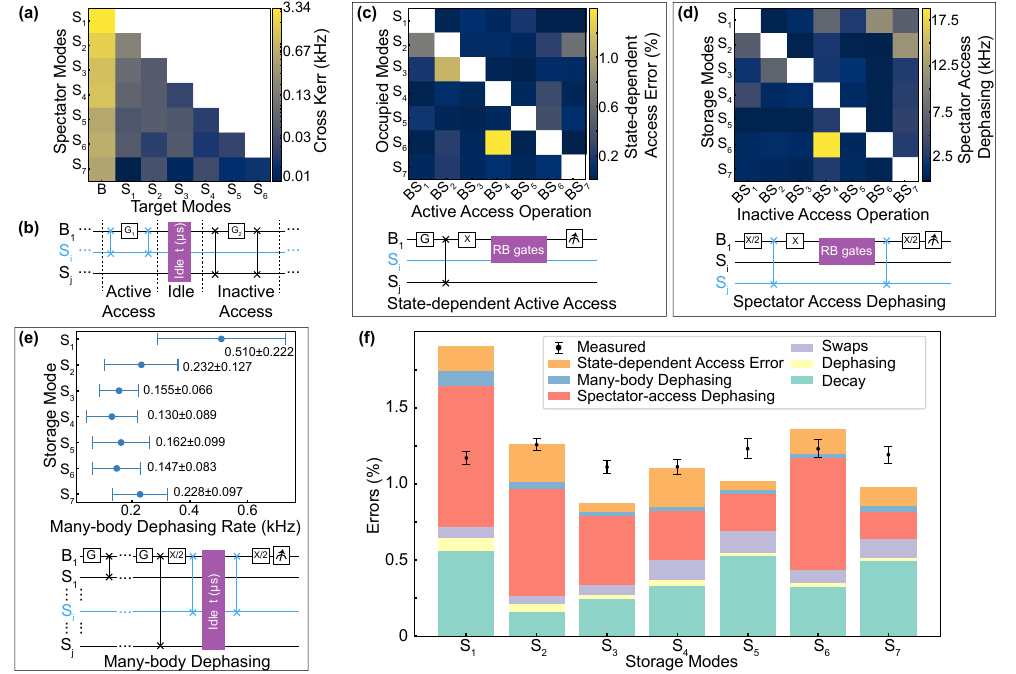}
    \caption{RAQM crosstalk benchmarking.  \textbf{(a)} Measured cross-Kerr values between storage and buffer modes (log-scale). \textbf{(b)} Three RAQM periods for a target storage mode $S_i$. Buffer gates $G_1$ and $G_2$ are applied during $S_i$'s active and inactive access periods. \textbf{(c)} $BS_i$ active access error due to states stored in $S_j$. The error rate is extracted using the bottom RB sequence. \textbf{(d)} The additional dephasing rate during $S_j$'s inactive access period, where the state stored in spectator modes $S_i$ \((i\neq j)\) undergoes random gates. This acts as a dephasing channel to $S_j$, with the rate measured using the bottom sequence.  \textbf{(e)} The many-body dephasing rate during the idle period, where the state stored in $S_i$ experiences additional dephasing due to cross-Kerr interactions between storage mode. This rate is measured using the bottom sequence. \textbf{(f)} The random read fidelity error budget. The dominant error channels come from storage decay and $BS_i$ cross-Kerr interactions. Details of the error budget are shown in Supplementary Section X.}
    \centering
    \label{fig:figure4}
\end{figure*}

\section{Exploring Many-body Interactions in Random Access Quantum Memory}

While bosonic modes in a linear cavity are naturally non-interacting, the introduction of nonlinear control elements induces parasitic cross-Kerr interactions between all modes (Fig.~\ref{fig:figure4}(a)). These interactions lead to state-dependent frequency shifts and many-body dephasing. To characterize the impact of these error channels, we divide the RAQM operation cycle into three distinct regimes with respect to a target storage mode $S_i$ (Fig.~\ref{fig:figure4}(b)). First, the \textbf{Active Access} period occurs when $S_i$ is actively swapped to or from the buffer. Second, the \textbf{Spectator Access} period covers the duration when another mode $S_{j \neq i}$ is being swapped, leaving $S_i$ idle but exposed to buffer activity. Finally, the \textbf{Idle} period corresponds to times when the buffer is empty, such as during processor computation. Crucially, Active errors occur a fixed number of times (twice) per cycle, while Spectator and Idle errors accumulate linearly with the memory size $N$.

During $S_i$'s active access period, the dominant crosstalk is the state-dependent access error (see Fig.~\ref{fig:figure4}(c)). Populations in other storage modes shift the $BS_i$ swap frequency by buffer-storage cross-Kerrs, reducing $S_i$'s access fidelity. To benchmark this error per active swap, we perform RB on $BS_i$ swap gates with different states stored in the other modes. Details about the RB fitting are shown in Supplementary Section VII. Overall, the \(BS_i\) gate incurs an additional error of $1.069\%$ due to populations in spectator modes.

During $S_i$'s inactive access period, the dominant crosstalk is the spectator access dephasing (see Fig.~\ref{fig:figure4}(d)). When a spectator mode ($S_{j}, j\neq i$) is accessed via $BS_j $ swap, the buffer mode acquires an unknown population, converting the buffer-storage cross-Kerr interactions into an effective dephasing channel. To extract this dephasing rate, we perform Ramsey experiments on $S_i$ while executing random gates between the buffer and each spectator mode \(S_j\). The average spectator access dephasing rate for storage pairs is $2.772\,\mathrm{kHz}$, which is slightly higher than storages' average Ramsey dephasing rate of \(2.105\,\mathrm{kHz}\).

During the idle period, the main crosstalk on $S_i$ is the storage many-body dephasing (see Fig.~\ref{fig:figure4}(e)). When all storage modes are occupied with unknown states, the storage-storage cross-Kerrs are converted into a many-body dephasing channel on $S_i$. To extract this error, we perform the Ramsey experiment on $S_i$ while storing identical cardinal states of the Bloch sphere in the other modes (See Supplementary Section VII for analysis). This storage many-body dephasing increases the storage Ramsey dephasing rate by $0.223\,\mathrm{kHz}$ on average.

The modes' decoherence and many-body interactions are the main error channels for memory operations. Fig.~\ref{fig:figure4}(f) shows the error budget of random read fidelity $\bar{F}_{BS_i}$ per gate of the size-7 RAQM. The state-dependent access error is determined by combining its swap infidelity from Section~\ref{sec: single mode memory access} with the state-dependent access error from its spectator modes. The decay, dephasing, and spectator-access dephasing errors for \(S_i\) are computed throughout its inactive access period. The many-body dephasing error is computed over the full duration of an RAQM cycle. Although the spectator-access dephasing error is dominant, it is due to the long inactive access period compared to other stages of RAQM control. Details about the error budget are shown in Supplementary Section X. 

Although the coupler and transmon have typical anharmonicities, the additional buffer layer shields the propagation of Kerrs, rendering the modes nearly transparent to each other. While many-body interactions are still the dominant error source, they are comparable to modes' decoherence~\cite{Chakram2022}. The dominant contribution to many-body interactions arises from stray cross-Kerr coupling to the buffer mode, which can be suppressed or even eliminated using a linearized coupler. The remaining many-body dephasing between storage modes increases weakly with memory size. However, it is expected to be small compared to individual mode decoherence for memory sizes of $10$–$10^2$ (Fig.~\ref{fig:figure4}(f)). At this scale, the RAQM can already achieve control-line efficiency (see Supplementary Section XII).

\section{Conclusion and Outlook}

We have realized a cascaded random access quantum memory. We demonstrate full multiplexed control of RAQM through one buffer mode and achieve random read fidelity beyond the memory error correction threshold. This enables the cascaded RAQM structure to serve as a building block for quantum error correction codes, using a single extra control line to enable scaling logical qubits by an amount limited by the coherence and the unwanted interactions among the memories.

While the fidelity demonstrated here is already above the nominal threshold for fault-tolerant memory, further improvements can enable practical use. The current dominant error, many-body interactions between buffer and storage modes, can be suppressed significantly using a more linearized coupler design with several junctions in series~\cite{zorin_flux_driven_twpa} and/or a balanced design~\cite{chapman_bs_2023, maiti2025linearquantumcouplerclean}. Niobium cavities~\cite{romanenko_cavities, milul_cavity_qubit, oriani2024niobiumcoaxialcavitiesinternal} have been proven to have $T_1$ beyond $20$ ms, which will significantly reduce the intrinsic decoherence error of the memories. Compact on-chip storage options using nanomechanical resonators~\cite{Cupertino2024} and spin memories~\cite{Reiner2024} can also provide longer coherence. With these improvements, the random read fidelity across all storage modes can exceed $99.9\%$ and support over $100$ memory modes (\(100\) times more logical qubits) compatible with the error correction, which can be randomly accessed and operated on with transversal two-qubit gates. The architecture is also compatible with hardware-efficient bosonic error correction codes~\cite{Ofek2016, marios_bosonic_QEC_2016, Campagne-Ibarcq2020, Sivak2023, Ni2023} and biased-noise/dual rail encodings~\cite{teoh_2023_dual_rail, Daul_rail_levine_2024, Chou2024_dual_rail, erasure_koottandavida_2024} that can further improve the amount of memory storage and overhead of quantum error correction. 

\section{Acknowledgement}

We thank Shannon Harvey, Fanghui Wan and Yueheng Shi for assistance with fabrication and cryogenic control. We are grateful to Connie Miao and Sho Uemura for their help with the FPGA setup and experiment code. We also thank Sebastien Leger for the insightful discussions on randomized benchmarking methods. This work was supported by AFOSR MURI (FA9550-19-1-0399, FA9550-21-1-0209, FA9550-23-1-0338). We acknowledge support from the Samsung Advanced Institute of Technology
Global Research Partnership. E.G. acknowledges support from NSF GRFP. The transmon chip was fabricated in the Pritzker Nanofabrication Facility at the University of Chicago, which receives support from the Soft and Hybrid Nanotechnology Experimental (SHyNE) Resource (NSF ECCS-1542205), a node of the National Science Foundation’s National Nanotechnology Coordinated Infrastructure. This work also made use of the shared facilities at the University of Chicago Materials Research Science and Engineering Center, supported by the National Science Foundation under award number DMR-2011854. The coupler chip was fabricated at the Stanford Nano Shared Facilities (SNSF) RRID: SCR023230, supported by the National Science Foundation under award ECCS-2026822. We gratefully acknowledge all the other members of the Schuster Lab for discussions and technical support.

\clearpage

\appendix

\section{Device Parameters}
\label{app:device_para}

\begin{table*}[t]
  \begin{tabular}{c c c c c c}
		    \hline
		    \hline
		    $\Phi_{\text{dc}}=0.269\pi$  & Symbol & Frequency/$2\pi$ (GHz) & $T_{1} \left(\mu \text{s}\right)$ & $T_{R} \left(\mu \text{s}\right)$ & $T_{\rm echo} \left(\mu \text{s}\right)$ \\  \hline
             Qubit ($Q$) & $\omega_{q}$ & $3.568$ & $493\pm13.1$ & $259\pm8.0$ & $370\pm14$  \\ \hline
             Coupler ($C$) & $\omega_{c}$ & $4.037$ & $45\pm0.8$ & $0.39\pm0.04$ &   \\ \hline
		     Buffer 1 & $\omega_{b1}$ & $4.984$ & $209\pm5.1$ & $75.5\pm3.8$ & $219\pm7.0$   \\ 
             Buffer 2 & $\omega_{b2}$ & $5.158$ & $120\pm1.3$ & $70.6\pm5.4$ & $193\pm6.7$  \\ \hline
             Storage 1 & $\omega_{s1}$ & $5.333$ & $358.3\pm6.0$ & $235.7\pm7.1$ & $551.6\pm16.8$  \\ 
             Storage 2 & $\omega_{s2}$ & $5.505$ & $1254.8\pm30.2$ & $378.3\pm15.1$ & $1493.2\pm53.5$  \\ 
             Storage 3 & $\omega_{s3}$ & $5.681$ & $799.0\pm12.7$ & $677.2\pm16.8$ & $1309.9\pm41.8$  \\
             Storage 4 & $\omega_{s4}$ & $5.860$ & $597.4\pm11.9$ & $806.2\pm23.7$ & $972.4\pm29.8$  \\
             Storage 5 & $\omega_{s5}$ & $6.037$ & $355.7\pm5.5$ & $591.8\pm15.2$ & $663.3\pm22.7$  \\
             Storage 6 & $\omega_{s6}$ & $6.229$ & $589.5\pm12.9$ & $1071.1\pm63.0$ & $1048.4\pm54.1$  \\
             Storage 7 & $\omega_{s7}$ & $6.407$ & $371.2\pm6.4$ & $663.5\pm29.0$ & $692.7\pm22.7$  \\
             \hline
             Dump 1 & $\omega_{d1}$ & $7.297$ & $0.73\pm0.03$ &  &   \\
             Dump 2 & $\omega_{d2}$ & $7.252$ & $1.81\pm0.11$ &  &   \\
             \hline
             Readout ($R$) & $\omega_{r}$ & $8.051$ & $0.05\pm0.01$ &  &   \\
             \hline\hline
             
		\end{tabular}
		\caption{Mode frequencies and coherence.}
		\label{table:frequency}
\end{table*}

\begin{figure}[t]
    \centering
    \includegraphics{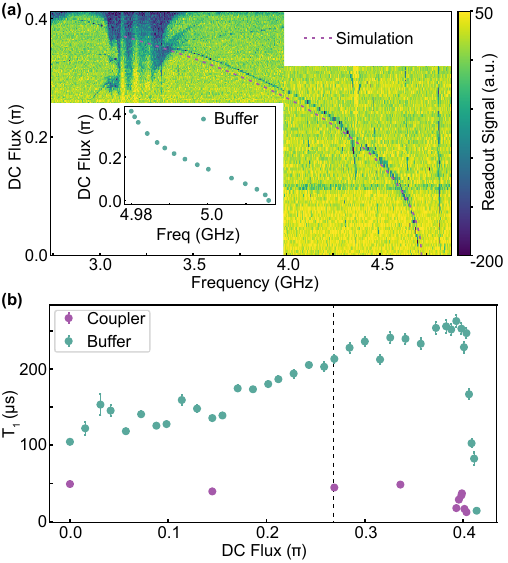}
    \caption{Coupler DC flux sweep. (a) Coupler and buffer (inset) mode frequencies. At each DC flux point, the buffer mode is initially prepared in $\ket{1}$, followed by an RF-flux modulation frequency sweep. The blue colors in the plot indicate possible transitions. The black-box simulation of the coupler frequency is shown as the red dashed line. The X-axis shows the coupler frequency by subtracting the buffer mode frequency. The scan is performed in two parts due to the RF channel frequency limitations. (b) T$_1$ measurements of the coupler and buffer modes. All RAQM experiments are conducted with the coupler at the dashed line ($0.269\pi$).}
    \centering
    \label{fig:t1sweep}
\end{figure}

\subsection{Coupler Dispersion}
We will first discuss the dispersion of the coupler and the buffer mode. Supplementary Figure~\ref{fig:t1sweep}(a) shows the coupler and buffer mode frequencies at different DC flux points, $\Phi_{\text{DC}}$. For each value of $\Phi_{\text{DC}}$, the buffer mode is first initialized to $\ket{1}$ using the transmon-buffer modes $\ket{f0} \leftrightarrow \ket{g1}$ sideband. The RF flux modulation is then turned on for $1 \, \mu\text{s}$, followed by another $\ket{f0} \leftrightarrow \ket{g1}$ sideband that maps the buffer state into the transmon $\{\ket{g}, \ket{f}\}$ subspace. This state is subsequently mapped to the $\{\ket{g}, \ket{e}\}$ subspace using a transmon $\pi$ pulse between $\ket{f}$ and $\ket{e}$. Blue regions in the plot indicate possible transitions between the buffer and other modes in the system. One of the most tunable traces corresponds to the coupler mode, which matches the black-box quantization results~\cite{nigg_bbox}.

The $T_1$ data for the coupler and buffer modes is shown in Supplementary Figure~\ref{fig:t1sweep}(b). To measure buffer mode's $T_1$, we prepare $|1\rangle$ in the buffer mode using the $|f0\rangle \leftrightarrow |g1\rangle$ sideband with the transmon, wait for a varied time, then apply the $|f0\rangle \leftrightarrow |g1\rangle$ sideband again and readout the qubit state. Since the coupler mode lacks a dedicated dispersive readout, we use a cascaded readout scheme: prepare $|1\rangle$ in $B$, apply an RF flux-modulated sideband between $|B\rangle$ and the coupler, transfer the excitation to the coupler, wait for a varied time, reverse the mapping, and finally read out the qubit state.

When the coupler is in the flux filter stopband (see Supplementary Section~\ref{app:filter}), the buffer mode's $T_1$ increases as $\Phi_{\text{DC}}$ increases. This behavior is attributed to the growing frequency difference between the coupler and the buffer mode, which reduces the dressing of the buffer mode. For $\Phi_{\text{DC}} > 0.4\pi$, the coupler frequency exits the filter stopband. Consequently, the buffer mode's $T_1$ drops significantly, and the coupler mode's $T_1$ becomes too short to be measured accurately with our FPGA electronics~\cite{Stefanazzi_2022_qick}.

\subsection{Device Coherences}
All modes' frequencies and coherences at the operation point $\Phi_{\text{DC}}=0.269\pi$ (dashed line in Supplementary Figure~\ref{fig:t1sweep}(b)) are shown in Supplementary Table~\ref{table:frequency}. In our experiments, the $T_1$ of the storage modes is limited by the external drive pin length, resulting in an over-coupled regime where external coupling $Q_c$ is smaller than the modes' intrinsic $Q_i$. We reduced the drive pin length in a subsequent cooldown to increase $Q_c$. In this new cooldown, we experimentally observed that $6$ out of the $7$ storage modes have $T_1$ exceeding $1\,\mathrm{ms}$. This coherence time is comparable to a bare Aluminum flute cavity~\cite{Chakram2022}. Further improvements would require different cavity materials, such as niobium~\cite{oriani2024niobiumcoaxialcavitiesinternal}. Detailed coherence time in the new cooldown is shown in Supplementary Table~\ref{table:coherence_new}.

\begin{table*}[t]
\begin{tabular}{c c c c c c c c}
\hline
\hline
 & Storage $1$ & Storage $2$ & Storage $3$ & Storage $4$ & Storage $5$ & Storage $6$ & Storage $7$ \\  \hline
 $T_1$ (ms) & $1.153\pm 0.020$ & $1.165\pm 0.019$ & $1.138\pm 0.016$ & $1.180\pm 0.009$ & $1.086\pm 0.016$ & $1.019\pm 0.014$ & $0.629\pm 0.007$ \\ \hline\hline
\end{tabular}
\caption{The experimentally measured storage mode $T_1$ in a different cooldown. The cavity drive pin length was reduced to prevent storage modes from being over-coupled. $S_7$ has a lower $T_1$ because of its proximity to the on-chip filter's stop band edge, resulting in reduced protection.}
\label{table:coherence_new}
\end{table*}

\subsection{Self-Kerrs and Cross-Kerrs}
Supplementary Table~\ref{table:kerrs} shows the cross-Kerrs and self-Kerrs measured in the experiments. The self-Kerrs \(K\) of the two buffer modes $B_i$'s (including the unused $B_2$ mode in the experiments) are measured through two coherent cavity displacements gapped by a varied waiting time $\tau$~\cite{Chakram2022}. The displacement amplitude $\alpha$ is also varied. After each sequence, the population of $B_i$ in the state $\ket{0}$ is measured, and the oscillation frequency, $f_d$, is extracted for each $\alpha$. Under small-angle approximation \(K|\alpha|^2t\ll \pi\) ~\cite{Chou2018}, the self-Kerr coefficient is determined as the fitted slope of $f_d$ to $\alpha$. $S_i$'s self-Kerrs are measured similarly, with the displaced state prepared in $B$ first and then swapped into $S_i$ through the $BS_i$ sideband. Cross-Kerrs between cavity modes are measured by performing Ramsey experiment on the target mode \(S_i\) with the spectator mode \(S_j\) initialized at $\ket{0}$ or $\ket{1}$. We measure the cross-Kerr strength between the coupler and $B_i(S_5)$ by performing cavity mode spectroscopy while the coupler is at $\ket{g}$ or $\ket{e}$. The cross-Kerrs between the coupler and other storage modes are measured through the $BS_i$ sideband frequency difference when the coupler is at $\ket{g}$ or $\ket{e}$.

\begin{table*}[t]
  \begin{tabular}{c c c c c}
		    \hline
		    \hline
		    Kerrs (kHz)  & $Q$ & $C$ & $B_1$ & $B_2$  \\  \hline
             $Q$ & $-143000$ &  &  &    \\ \hline
             $C$ & - & $-57120$ &  &    \\ \hline
		     $B_1$ & $-285$ & $-1452$  & $-6.8$  &     \\ 
             $B_2$ & $-271$ & -  & $-15$  & $-5.1$   \\ \hline
             $S_1$ & - & $-422$ & $-3.336\pm0.154$ & -  \\ 
             $S_2$ & - & $-412$  & $-1.838\pm0.171$ & -  \\ 
             $S_3$ & - & $-202$ & $-1.192\pm0.167$ & -  \\
             $S_4$ & - & $-62$ & $-1.344\pm0.173$ & -  \\
             $S_5$ & - & $-278$ & $-0.600\pm0.163$ & -  \\
             $S_6$ & - & $-192$ & $-0.716\pm0.172$ & -  \\
             $S_7$ & - & $-232$ & $-0.525\pm0.172$ & -   \\
             \hline
             $R$ & $-320$ & - & - & -   \\
             \hline\hline
             
		\end{tabular}
        \begin{tabular}{c c c c c c c c}
		    \hline
		    \hline
		    Kerrs (kHz)  & $S_1$ & $S_2$ & $S_3$ & $S_4$ & $S_5$ & $S_6$ & $S_7$ \\  \hline
             $S_1$ & $-0.841$ &  &  &  &  &  &   \\ 
             $S_2$ & $-0.178\pm0.011$ & $-0.062$ &  &  &  &  &   \\ 
             $S_3$ & $-0.076\pm0.009$ & $-0.044\pm0.01$ & $0.073$ &  &  &  &   \\
             $S_4$ & $-0.061\pm0.009$ & $-0.046\pm0.009$ & $-0.020\pm0.009$ & $0.954$ &  &  &   \\
             $S_5$ & $-0.037\pm0.01$ & $-0.046\pm0.009$ & $-0.038\pm0.009$ & $-0.016\pm0.009$ & $-0.466$ &  &   \\
             $S_6$ & $-0.043\pm0.014$ & $-0.017\pm0.014$ & $-0.041\pm0.016$ & $-0.012\pm0.014$ & $-0.019\pm0.015$ & $-0.194$ &   \\
             $S_7$ & $-0.005\pm0.010$ & $-0.030\pm0.009$ & $-0.007\pm0.010$ & $-0.025\pm0.009$ & $-0.009\pm0.008$ & $-0.011\pm0.010$ & $0.048$   \\
             \hline\hline
             
		\end{tabular}
		\caption{Mode self-Kerrs and cross-Kerrs at $\Phi_{\text{dc}}=0.269\pi$. Positive Kerrs might come from hybridizing to the coupler's higher differential mode.}
		\label{table:kerrs}
\end{table*}

\section{System Hamiltonian}
\label{app:system_hamiltonian}
\begin{figure}[t]
    \centering
    \includegraphics{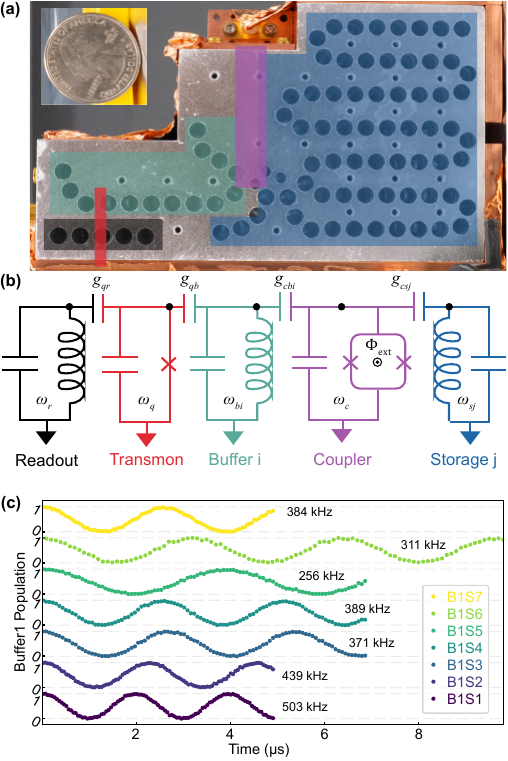}
    \caption{(a) False-color device image. (b) RAQM circuit diagram. For simplicity, only the couplings to the buffer mode $B_i$ and the storage mode $S_j$ are shown. The coupler is capacitively coupled to all buffer, storage, and dump modes. (c) Experimentally measured $B_1S_j$ sideband rates.}
    \centering
    \label{fig:supp_circuits}
\end{figure}

Supplementary Figure~\ref{fig:supp_circuits}(a) provides a physical illustration of the RAQM. The dump modes are technically unprotected storage modes at higher frequencies that occupy the same locations as the storage modes. The other RAQM components are color-coded according to their physical placement. 

Supplementary Figure~\ref{fig:supp_circuits}(b) shows the RAQM circuit diagram. The effective capacitances and inductances of the different modes are calculated using Black-box quantization~\cite{nigg_bbox} from the admittance extracted via HFSS simulations. To calculate the memory sideband rate analytically, we first consider the subsystem consisting of $B_i$, the coupler, and $S_j$ for simplicity. The corresponding dressed frequencies are $\omega_{b_i}$, $\omega_c$, and $\omega_{s_j}$. Assuming symmetric Josephson junctions with energy \(E_J\), and a purely differential RF flux drive~\cite{luSchoelkopf2023}, the potential energy of the coupler SQUID is
\begin{align}\label{eq:squid_dds}
U = 2E_{J}\cos\!\left(\varphi_{\text{ext}}\right)\cos\!\left(\varphi_c \right).
\end{align}
Here $\varphi_{\text{ext}} = \pi\Phi_{\text{ext}}/\Phi_0$ is the external flux threading the SQUID loop, and $\varphi_c$ is the phase variable associated with the coupler mode in the dressed basis. Under an external RF flux modulation of amplitude \(\varepsilon\) and frequency \(\omega_d\), together with a DC flux bias \(\varphi_{\text{DC}}\), we have
\begin{align} \label{eq:phi_ext}
\varphi_{\text{ext}} = \varphi_{\text{DC}}+\varepsilon\sin(\omega_d t).
\end{align}

Assuming \(\varepsilon\ll 1\) (\(\varepsilon = 0.035\) from simulations), we substitute Eq.~\ref{eq:phi_ext} into Eq.~\ref{eq:squid_dds} and expand to first order in \(\varepsilon\):
\begin{align}
U \approx 2E_{J}\Bigl(\cos\varphi_{\text{DC}} - \varepsilon\sin(\omega_d t)\sin\varphi_{\text{DC}} \Bigr)\cos\varphi_c .
\label{eq:potential2}
\end{align}

When the coupler is dispersively coupled to the other modes, in the appropriate rotating frame, $\varphi_c$ can be approximated by annihilation operators:
\begin{align}
\varphi_c &= \varphi_{\text{zpf}}\Bigl(a_ce^{i\omega_ct}-\frac{g_{c b_i}}{\Delta_{c b_i}}a_{b_i}e^{i\omega_{b_i}t} \nonumber \\
&\hspace{2cm} -\frac{g_{c s_j}}{\Delta_{c s_j}}a_{s_j}e^{i\omega_{s_j}t}+\text{h.c.}\Bigr),
\label{eq:expansion}
\end{align}
where \(\varphi_{\text{zpf}}\) is the zero-point fluctuation of the phase variable. The dressed operators for the coupler mode include contributions from each mode, normalized by participation factors of the form \(g/\Delta\), where \(g_{c x}\) is the coupling strength and \(\Delta_{c x} = \omega_c - \omega_x\) is the detuning between the coupler and mode \(x\). 

Plugging Eq.~\ref{eq:expansion} into Eq.~\ref{eq:potential2} and expanding the potential up to second order in the participation factors, we choose $\omega_d = \omega_{b_i}-\omega_{s_j}$ to activate the beamsplitter interactions. Keeping only the time-independent terms, we obtain
\begin{align} \label{eq:swap_rate}
U_{\text{swap}} \approx \underbrace{\varepsilon E_{J}\sin(\varphi_{\text{DC}}) \varphi_{\text{zpf}}^2\frac{g_{c b_i}}{\Delta_{c b_i}}\frac{g_{c s_j}}{\Delta_{c s_j}}}_{g_{\textrm{swap}, b_i s_j}}\bigl(a_{b_i}a_{s_j}^{\dagger}+\text{h.c.}\bigr),
\end{align}
where \(g_{\textrm{swap}, b_i s_j}\) denotes the beamsplitter rate. In experiments, we calibrate this beamsplitter gate by preparing a photon in the buffer mode and activating the beamsplitter drive for \(\sim100\,\mu\text{s}\) (equivalent to \(\sim50\) swaps) across a range of frequencies. We then choose the resonant frequency that maximizes the probability of finding the photon in the storage mode. The repeated sequence of swap gates helps calibrate the pulse parameters in the presence of coherent errors. For calibration under bichromatic drive tones or strong Stark shifts, we refer the reader to Ref.~\cite{luSchoelkopf2023}.

Supplementary Figure~\ref{fig:supp_circuits}(c) shows the sideband rates used in the RAQM experiments. The primary limitation to our gate speeds are frequency collisions between these sidebands. For example, the \(B_1S_1\) (\(\omega_{s_1} - \omega_{b_1} = 2\pi \times 349\,\text{MHz}\)) and the \(S_1S_3\) (\(\omega_{s_3} - \omega_{s_1} = 2\pi \times 348\,\text{MHz}\)) differ by only \(1\,\text{MHz}\). The size of collisions for other \(B_1S_j\) sidebands ranges from \(2{-}20\,\text{MHz}\). These frequency collisions limit our gate speeds to \(300{-}500\,\text{kHz}\). Such issues can be avoided by choosing an appropriate frequency for the buffer mode; since the frequency gaps between storage modes are roughly equal, the buffer mode should ideally be half the gap away from the first storage mode: \(\omega_{s_1} - \omega_{b_1} \approx (\omega_{s_2} - \omega_{s_1})/2\).

Wheras the swap interaction comes from terms second order in \(\varphi_c\), the cross-Kerr/self-Kerr interaction comes from the fourth order terms. For instance, mode \(B_i\)'s self-Kerr \(k_{b_i}\) of mode  can be derived as  
\begin{align}
    U_{\text{kerr}} &\propto  \frac{\partial^4 U}{\partial \varphi_c^4}\Bigr|_{\varphi_c=0}\left(\varphi_{\text{zpf}}\frac{g_{cb_i}}{\Delta_{cb_i}}\right)^4 (a_{cb_i}^\dagger a_{cb_i})^2 \\
    &\propto \underbrace{ (2E_J \cos{\varphi_{DC}})\left(\varphi_{\text{zpf}}\frac{g_{cb_i}}{\Delta_{cb_i}}\right)^4 }_{k_{b_i}}(a_{cb_i}^\dagger a_{cb_i})^2 
\end{align}
up to a constant factor. Note that while the current discussion is limited to the coupler, the qubit also contributes to the self-kerr of the buffer modes. The cross-Kerr $\chi_{b_is_j}$ between buffer \(B_i\) and storage \(S_iS_j\) modes can be derived in a similar way using an additional identity that the cross-Kerr is a geometric mean of the modes' self-Kerrs: 
\begin{align} \label{eq: Cross_kerr}
   \chi_{b_is_j} &= -2\sqrt{k_{b_i}k_{s_j}}.
\end{align}
The ratio then between the undesired cross-Kerr from Eq.~\ref{eq: Cross_kerr} and the desired swap interaction from Eq.~\ref{eq:swap_rate} is 
\begin{align}
    \frac{\chi_{b_is_j}}{g_{\text{swap}, b_is_j}} \propto \frac{\varphi_{\text{zpf}}^2 }{\epsilon \tan{\varphi_{DC}}}\frac{g_{cb_i}}{\Delta_{cb_i}}\frac{g_{cs_j}}{\Delta_{cs_j}}
\end{align}
up to a constant factor. This ratio can be optimized by reducing the zero-point phase fluctuations through arraying larger junctions in series \cite{zorin_flux_driven_twpa}. In addition, there are other coupler designs such as LINC \cite{maiti2025linearquantumcouplerclean} and SNAIL \cite{chapman_bs_2023} that completely eliminate the idle cross-Kerr between modes.

The cross-Kerr between buffer mode $B_i$ and storage mode $S_j$, and the modes' self-Kerrs can also be expressed in terms of effective mode capacitance $\{C_{b_i}, C_{s_j}\}$ and inductance $\{L_{b_i}, L_{s_j}\}$ as ~\cite{nigg_bbox}:
\begin{align}
k_{b_i(s_j)} &= \frac{L_{b_i(s_j)}}{C_{b_i(s_j)}}\frac{C_{c}}{L_{c}}E_{C_c}, \\
L_c &= \frac{\Phi_0^2}{4\pi^2E_{Jc}\cos\left(\varphi_{\text{DC}}\right)}
\end{align}
Here, $E_{C_c}$ is the charging energy for the coupler, and $C_c$ is the coupler's capacitance to ground.

\section{Active Reset}
\label{app:active}

The long coherence of the storage modes results in an extended system reset time ($5T_1>5$ ms). To speed up all the experiments, we implement an active reset protocol modularly, including buffer mode reset, qubit reset, and storage mode reset. 

\begin{figure}[t]
    \centering
    \includegraphics{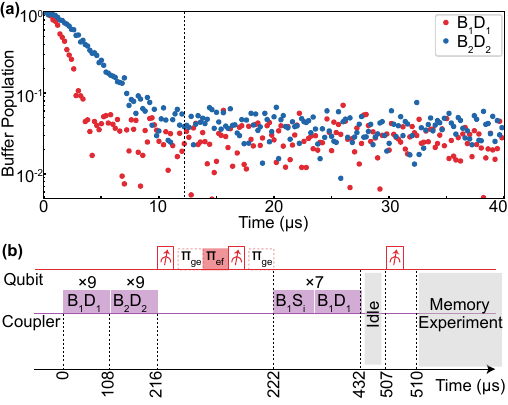}
    \caption{RAQM active reset. (a) Buffer-Dump modes swap. Dash line ($12 \mu$s) shows the time experiments used to evacuate buffer modes. (b) Active reset sequence. The sequence starts with evacuating buffer modes. Multiple buffer-dump swaps at different frequencies (dressed by $\chi_{qb_i}$) are sequentially applied. Qubit state (up to $\ket{f}$ level) is actively reset. All Storage modes' populations are sequentially swapped into $B_1$, then evacuated through $D_1$. After $75\,\mu\mathrm{s}$' idling period, the experiment starts when the qubit state is $\ket{g}$.}
    \centering
    \label{fig:activereset}
\end{figure}

Supplementary Figure~\ref{fig:activereset} (a) illustrates the sideband interactions between the buffer and dump modes. We separately prepare $\ket{1}$ in $B_1$ and $B_2$, then activate the coupler modulation at the difference frequencies of $B_1D_1$ and $B_2D_2$ for a variable time. The lossy dump modes, acting as cold reservoirs, autonomously dissipate the received excitations, effectively cooling the buffer modes. The residual population in the buffer modes is swapped back to the qubit's $\ket{f}$ level, rotated to $\ket{e}$ level, and subsequently measured. The residual population of both buffer modes  remains constant at \(<10\%\) after $12\,\mu\mathrm{s}$ of sideband interactions.

Supplementary Figure~\ref{fig:activereset} (b) illustrates our RAQM active reset protocol, performed before the main experiments. The buffer modes are evacuated first. In the memory experiments, both the buffer modes and the qubit theoretically use, at most, the first excited states. Consequently, the protocol ensures state reset upto the second excitation for all modes. Since the qubit-$B_i$ dispersive shift $\chi_{qb_i}$ is comparable to the sideband rate, we sequentially choose $9$ different coupler RF modulations frequencies for each $B_iD_i$ sideband, accounting for the dressing effects of the dispersive shifts. After evacuating both buffer modes, we sequentially reset the qubit's $\ket{e}$ and $\ket{f}$ states. First, we measure the qubit state using a $1\,\mu\mathrm{s}$ readout pulse, wait for $2\,\mu\mathrm{s}$, and conditionally flip the qubit state if it is in the $\ket{e}$ state. Next, we apply a qubit $\ket{f}$-$\ket{e}$ $\pi$ pulse and repeat the active reset for the $\ket{e}$ state. To reset the storage modes, we sequentially apply $7$ swap pairs, transferring the population from $S_i$ to $B_1$, then dump the population from $B_1$ to $D_1$ and autonomously evacuated. Finally, before starting the memory experiments, we idle the system for $75\,\mu\mathrm{s}$ so that any remaining low-coherence modes are fully reset. The experiments are initiated only when the qubit is confirmed in the $\ket{g}$. The qubit readout performance in our experiments is shown in Supplementary Table~\ref{table:readout}.

\begin{table}[t]
  \begin{tabular}{c c c }

            \hline
		    \hline
		      & Symbol & Value  \\  \hline\hline
             Readout length & $t_r$ & $1.0\,\mu\text{s}$  \\ \hline
             Readout relaxation time & $t_{gr}$ & $2.5\,\mu\text{s}$  \\ \hline
		     Assignment fidelity ($\ket{g}$) & $P_{gg}$ & $99.68\%$ \\ \hline
             Assignment fidelity ($\ket{e}$) & $P_{ee}$ & $98.34\%$ \\ \hline\hline
             
		\end{tabular}
		\caption{Qubit readout performance. The readout fidelity is estimated by averaging $20000$ single-shot histograms.}
		\label{table:readout}
\end{table}

\section{System Temperature}
\label{app:temp}
\begin{figure}[t]
    \centering
    \includegraphics{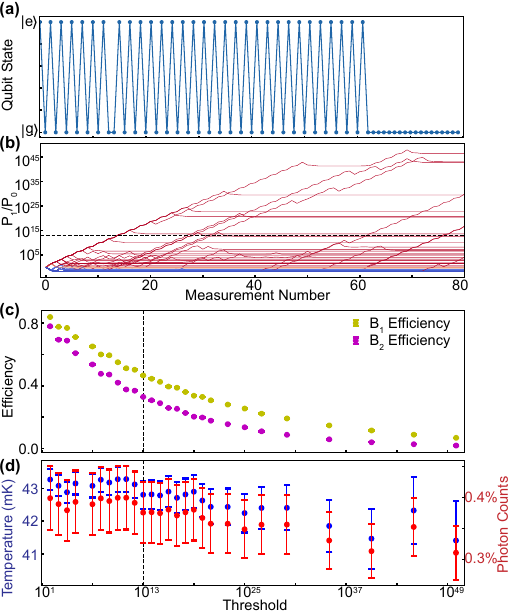}
    \caption{Cavity temperature measurement. (a) The oscillatory readout signal indicates the presence of a photon in the cavity by continuously monitoring the cavity parity. (b) Each readout signal string is mapped to the cavity state probability through hidden Markov analysis. Red traces show the ratio change for different readout signals as strings are analyzed. The analyzed detection efficiency (c) and temperature (d) depend on the threshold. Here, we show the temperature for $B_1$ as an example. $2 \times 10^5$ measurements are used to approximate the detection efficiency and temperature.}
    \centering
    \label{fig:temp_measurement}
\end{figure}

We benchmark the cavity's temperature using the Ramsey parity measurement~\cite{dixit_dark_matter}. To detect the thermal photon population in $B_1$, we repeatedly measure the parity information 80 times in a single experiment. Each measurement sequence consists of the following pulses: a transmon $\pi/2$ rotation, an idle period of $\pi/\chi$, another transmon $\pi/2$ rotation, a readout pulse ($1\,\mu\mathrm{s}$), and a readout evacuation period ($2.5\,\mu\mathrm{s}$). Each experiment produces an 80-bit string indicating the transmon's state. Supplementary Figure~\ref{fig:temp_measurement}(a) illustrates the string behavior when $B_1$ is initialized to $\ket{1}$.

Given a specific type of transmon state string, we quantify whether the cavity is populated by applying the hidden Markov process~\cite{dixit_dark_matter} to calculate the cavity population ratio:$\frac{P_1}{P_0}$. Here, the cavity state probabilities $P_0$ and $P_1$ correspond to the cavity being in $\ket{0}$ and $\ket{1}$, respectively. These probabilities are reconstructed using the backward propagation algorithm based on the transmon state string. Starting from the beginning of the state string, the algorithm iteratively calculates the likelihood of observing the measured readout signal up to the current position, assuming the cavity was initially in $\ket{0}$ or $\ket{1}$. Each transmon state string is then converted into a probability string.

In Supplementary Figure~\ref{fig:temp_measurement}(b), we analyzed $10^5$ measurements taken when the system is in its thermal equilibrium state. A higher ratio in the plot indicates a greater likelihood of photon population in the buffer modes. As the threshold increases (dashed line), measurement errors are filtered out, isolating events corresponding to thermal photons.

After extracting the thermal photon events, we need to compensate for two key factors to avoid overestimating the system temperature: detection efficiency and event assignment. The detection efficiency $\eta_i$ describes the probability of detecting a photon when $B_i$ is in $\ket{1}$. In our experiments, we separately prepare $\ket{1}$ in $B_1$ and $B_2$, perform the Ramsey parity measurement to count cavity photon events, and approximate $\eta_i$ using $2 \times 10^5$ experiments. In Supplementary Figure~\ref{fig:temp_measurement}(c), fewer photon events are captured as the threshold increases. Since both $B_1$ and $B_2$ are coupled to the transmon, thermal population in either cavity can trigger the detection. We assume that both modes have the same temperature $T_c$. Given the experimentally measured total photon counts $\delta$ and efficiency $\eta_i$, the temperature $T_c$ required to explain the counts should satisfy:

\begin{align}
\sum_{i=1,2}\frac{\eta_i}{-1+\exp{(\frac{\hbar\omega_{bi}}{k_BT_c}})}=\delta\
\end{align}

Supplementary Figure~\ref{fig:temp_measurement}(d) shows the corrected thermal photon counts in $B_1$ and the corresponding temperature. We choose $10^{13}$ as our threshold ($\eta_1=46.6\%\pm0.2\%$, $\eta_2=33.0\%\pm0.2\%$) and list the system temperature (steady state temperature $T_{SS}$ and photon counts $\bar{n}_{ss}$, active reset temperature $T_{ar}$ and photon counts $\bar{n}_{ar}$ in Supplementary Table~\ref{table:temperature}. Qubit temperature is calculated by measuring $\ket{e}\leftrightarrow\ket{f}$ Rabi oscillation amplitude with and without initial $\pi$ pulse between $\ket{g}$ and $\ket{e}$. The temperature of the other storage and dump modes is measured with an additional $\pi$ swap between corresponding buffer modes initially. The coupler's temperature is higher after the active reset, which can be mitigated by adding additional coupler reset through the dump modes or waiting longer ($>75\,\mu$s) before the experiments start.

\begin{table}[t]
  \begin{tabular}{c c c c c}
		    \hline
		    \hline
		    $\Phi_{\text{dc}}=0.269\pi$ &  $\bar{n}_{ss} (\%)$ & $T_{ss}$ (mK) & $\bar{n}_{ar} (\%)$ & $T_{ar}$ (mK) \\  \hline
             Qubit & $1.59\pm0.12$ & $41.2\pm0.01$ & - & -  \\ 
             Coupler & $1.18\pm0.12$ & $43.5\pm0.3$ & $4.96\pm0.25$ & $63.5\pm0.3$   \\ \hline
             Buffer 1 & $0.38\pm0.05$ & $42.8\pm0.4$ & $0.55\pm0.06$ & $45.8\pm0.4$   \\
             Buffer 2$^*$ & - & $42.8\pm0.4$ & - & $45.8\pm0.4$  \\
             \hline
             Storage 1 & $0.20\pm0.03$ & $41.3\pm0.5$ & $0.59\pm0.06$ & $49.9\pm0.4$  \\ 
             Storage 2 & $0.14\pm0.02$ & $40.4\pm0.5$ & $0.48\pm0.06$ &  $49.4\pm0.4$ \\ 
             Storage 3 & $0.12\pm0.02$ & $40.8\pm0.5$ & $0.33\pm0.04$ &  $47.8\pm0.4$ \\ 
             Storage 4 & $0.11\pm0.02$ & $41.4\pm0.5$ & $0.27\pm0.03$ &  $47.5\pm0.5$ \\ 
             Storage 5 & $0.08\pm0.02$ & $40.6\pm0.5$ & $0.24\pm0.03$ &$48.1\pm0.2$   \\ 
             Storage 6 & $0.17\pm0.02$ & $46.7\pm0.5$ & $0.61\pm0.05$ &  $58.5\pm0.2$ \\ 
             Storage 7 & $0.05\pm0.01$ & $40.9\pm0.6$ & $0.17\pm0.02$ & $48.1\pm0.2$  \\
             \hline
             Dump 1 & $0.02\pm0.004$ & $40.8\pm0.7$ & $0.03\pm0.005$ & $42.8\pm0.3$  \\ 
             Dump 2 & $0.02\pm0.004$ & $41.4\pm0.6$ & $0.04\pm0.008$ & $45.1\pm0.3$  \\ 
             \hline\hline
		\end{tabular}
		\caption{System temperature and thermal photon populations. \\
                $^*$ Buffer 2's temperature is assumed to be the same as Buffer 1.}
		\label{table:temperature}
\end{table}

\section{Phase Calibration for RAQM experiments}
\label{sec:phase calibration}

\begin{figure}[t]
    \centering
    \includegraphics{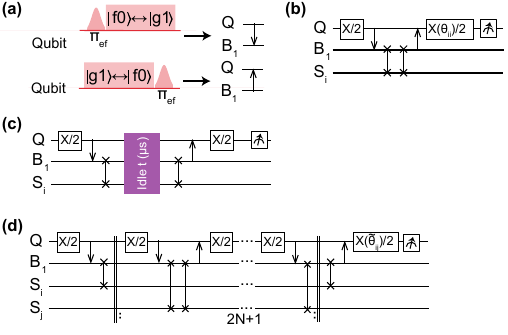}
    \caption{RAQM Phase calibration. Each transmon gate's virtual-$Z$ phase comprises three components: the active-access phase $\theta_{ii}$, the inactive-access phase $\theta_{ij}$ accumulated during storage read and write, and the time-dependent idle phase $\theta_{\Delta_i}(t)$ caused by hardware frequency tracking. (a) Pulse sequences for state transfer between transmon and $B_1$. (b) active-access phase $\theta_{ii}$ calibration for $S_i$. (c) Idle phase calibration extracts the Stark-shift $\Delta_i$ to $S_i$ through the Ramsey-like experiments. (d) Inactive-access phase $\theta_{ij}$ calibration on $S_i$ after $S_j$ access operations. $\theta_{ij}$ are calculated based on the maximum contrast phase $\tilde{\theta}_ij$ and $\Delta_i$. To improve calibration accuracy, $N+1$ writes and $N$ reads are repeated. }
    \centering
    \label{fig:phase_calibration}
\end{figure}

In the RAQM experiments, states are modified on the transmon side and transferred to $B_1$ and $S_i$. However, the beamsplitter interactions between buffer and storage modes naturally provide iSWAP instead of the SWAP gate ($\pi/2$ phase difference). The sideband interactions used for state transfer also induce Stark shifts in mode frequencies. As a deterministic frequency shift, this phase accumulation can all be corrected by adjusting the phases of all subsequent transmon gates~\cite{PhysRevResearch.2.033447}, effectively acting as an additional virtual-Z gate.

The virtual-$Z$ phase is the sum of the following three components:  

(a) Active-access phase ($\theta_{ii}$): This is the phase accrued in $S_i$ during its read and write operations.  

(b) Inactive-access phase ($\theta_{ij}$): This is the phase accrued in $S_i$ during the read and write operations of $S_j$.  

(c) Idle phase ($\theta_{\Delta_i}(t)$): When sidebands ($Q \leftrightarrow B_1$, $B_1 \leftrightarrow S_i$) are pulsed, the FPGA board tracks the Stark-shifted frequency $f_{ss}$. During the periods when the sidebands are off, the system’s idle frequency $f_{id}$ differs from the tracked frequency, causing all subsequent sideband pulse phases to deviate by $\Delta_{i}t_{\text{gap}}$. Here $\Delta_{i}=f_{ss} - f_{id}$ is the total Stark-shift of $S_i$ due to sidebands.

Supplementary Figure~\ref{fig:phase_calibration} shows the process of virtual-$Z$ phase calibration. For simplicity, state read and write operations between the transmon and $B_1$ are represented using circuit symbols in Supplementary Figure~\ref{fig:phase_calibration}(a). Supplementary Figure~\ref{fig:phase_calibration}(b) shows the circuit to calibrate $\theta_{ii}$. The phase of the second transmon gate is swept, and $\theta_{ii}$ is determined as the phase that maximizes the final $\ket{e}$ population. Supplementary Figure~\ref{fig:phase_calibration}(c) shows the circuit to calibrate $\theta_{ii}$. By sweeping the waiting time between two $B_1S_i$ swaps, we fit the Ramsey signal and extract the Stark-shift frequency $\Delta_i$. Supplementary Figure~\ref{fig:phase_calibration}(d) shows the circuit to calibrate $\theta_{ij}$. Since the inactive-access phase is small, we repeat the read and write operations on $S_j$ for $5$ times. The maximum contrast phase $\tilde{\theta}_{ij}$, determined by the maximum final $\ket{e}$ population in the transmon, can be used to calculate $\theta_{ij}$: 
\begin{align}
\theta_{ij}=\frac{\tilde{\theta}_{ij}-\Delta_{i}t_{\text{gap}}}{5}
\end{align}
Here, $t_{\text{gap}}$ is the time between the first and the last $B_1S_i$ swaps.
For the multiplexed RAM RB experiments, each $S_i$ tracks its virtual-$Z$ phase based on the RB sequences. The phases of all transmon gates are first calculated on a classical computer and then sent as instructions to the FPGA board to generate the corresponding pulses.

\section{Fast Flux Delivery for High-Q 3D Cavities}
\label{app:filter}
\begin{figure}[t]
    \centering
    \includegraphics{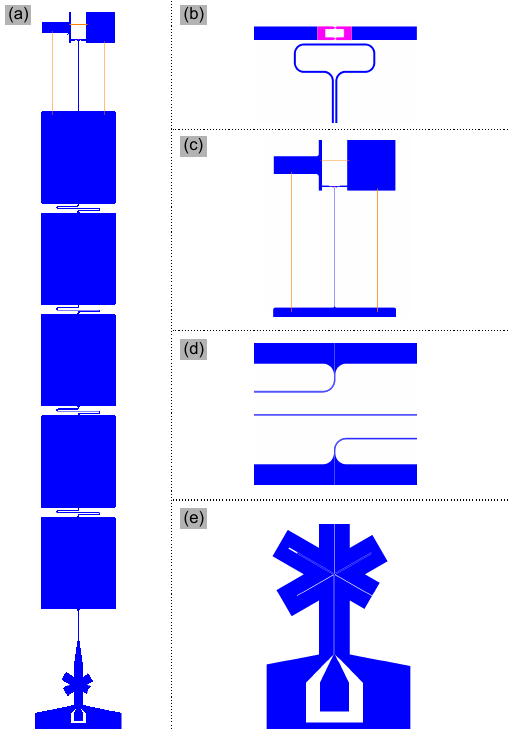}
    \caption{Fast flux delivery for 3D cavities. (a) Overview of the coupler chip, which comprises four main components from top to bottom: (b) the SQUID coupler with its associated flux loop, (c) shorting line structures (in orange) for ESD protection, (d) the edge-coupled microstrip filter, and (e) the double-Y balun. The blue areas indicate Nb metal regions, while pink highlights the Al SQUID junctions. The chip is inserted into the multimode cavity and wire-bonded to an SMA connector for DC and RF flux control.}
    \centering
    \label{fig:coupler_design}
\end{figure}

Flux control is essential in quantum computing for implementing high-fidelity two-qubit gate schemes and single transmon Z-control. While on-chip flux lines are commonly used in 2D systems, integrating flux lines in 3D cavities is challenging due to two primary factors:

(1) The absence of a well-defined ground plane on the chip.

(2) The widely distributed cavity mode field, which can easily couple to the flux line, substantially reduces the cavity $T_1$ without flux line protection.

In 3D systems, introducing flux lines while preserving cavity coherence remains a significant challenge. For DC flux tuning, external flux coils are commonly used, but their reliance on normal metals limits cavity coherence. Recently, on-chip flux pick-up loops~\cite{chapman_bs_2023} and magnetic hoses~\cite{Valadares2024} have emerged as promising alternatives for DC flux biasing. In contrast, RF flux modulation has only been demonstrated using a coaxial stub cavity~\cite{luSchoelkopf2023}. Despite these advancements, an on-chip flux line capable of supporting both DC and RF signals while maintaining high coherence in a single cavity mode has yet to be realized.

Here, we realized a simple on-chip flux line design compatible with standard 3D cavities for both DC and RF flux delivery, with the following advantages:

(1) Ultra-wide ($>3$ GHz) and deep stopband with a flexible flux line geometry.

(2) Compatible with coaxial connectors, one of the most widely used types for RF experiments.

(3) Requiring only a single standard superconducting chip, eliminating the need for additional circuit components.

(4) Free from external flux coils while maintaining high cavity coherence ($>1$ ms) and low cavity thermal populations ($<0.05\%$).

Our coupler chip design is shown in Supplementary Figure~\ref{fig:coupler_design}(a). It is composed of 4 parts (shown in Supplementary Figure~\ref{fig:coupler_design}(b), (c), (d), (e)): The SQUID coupler, the shorting lines, the edge-coupled microstrip filter, and the double-Y balun. 

The SQUID coupler is a flux-tunable transmon with two pads that are capacitively coupled to both buffer and storage modes. The flux loop is designed to deliver a pure RF flux modulation to the SQUID. However, the flux loop also has a stray capacitive coupling to the coupler. One solution is to maximize the flux line's inductive coupling to the SQUID loop~\cite{Li2024, Li2024_autonomous}. In these references involving 2D designs, the flux line produced by such optimization is effective for all frequency ranges and helps mediate fast and clean two-mode squeezing sidebands at $>7\,\mathrm{GHz}$. In this work, we can alternatively optimize the SQUID coupler geometries for low RF modulation frequencies $<2\,\mathrm{GHz}$~\cite{luSchoelkopf2023, Lu2025}.
We optimized the location of the flux loop using ANSYS HFSS simulations. In particular, we quantified the flux-to-charge drive ratio with the following equations: We define the two SQUID junctions as $J_1$ and $J_2$. When RF signals at frequency $\omega_{rf}$ are applied through the flux line, the induced currents along $J_1$ and $J_2$ are denoted $I_1(\omega_{rf})$ and $I_2(\omega_{rf})$, respectively~\cite{luSchoelkopf2023}. For a pure flux drive:
\begin{equation}
I_1(\omega_{rf}) = -I_2(\omega_{rf})
\end{equation}
For a pure charge drive:
\begin{equation}
I_1(\omega_{rf}) = I_2(\omega_{rf})
\end{equation}
In the presence of both drives, maximizing the flux modulation amplitude is equivalent to minimizing the following ratio $r(\omega_{rf})$:
\begin{equation}
r(\omega_{rf}) = \left| \frac{I_1(\omega_{rf}) + I_2(\omega_{rf})}{I_1(\omega_{rf}) - I_2(\omega_{rf})} \right|
\end{equation}
Following ref.~\cite{luSchoelkopf2023}, we optimize the flux-loop position and capacitor pad geometry to minimize $r(\omega_{rf})$. In practice, optimizing across the entire range of $\omega_{rf}$ is challenging. Since only RF modulation frequencies below $2.5$ GHz are required, we focused our HFSS optimization on the low-frequency range.

Shorting lines are for ESD protection of fab on sapphire chips. They are removed later during packaging and are discussed in the Supplementary Section.~\ref{app:ESD}.

The edge-coupled microstrip filter~\cite{zhao2025fluxtunablecavitydarkmatter} provides a wide stopband to protect all coherent elements in our RAQM. As shown in Supplementary Figure~\ref{fig:filter_S21}(c), the simulated response (blue curve) demonstrates this stopband. Similar to a Step-Impedance Low-Pass Filter (SIPF), the edge-coupled microstrip filter consists of alternating high- and low-impedance coplanar stripline (CPS) sections of equal electrical length. The stopband’s width, depth, and center frequency are determined independently by the impedance ratio, the number of sections, and the electrical length. The analytic expression~\cite{filter_IEEE} can be used to quickly calculate the ABCD matrix of the filter and approximate the stopband shape.

The double-Y balun~\cite{DoubleYbalun_1992, balun_IEEE} is used for connecting a balanced circuit (edge-coupled microstrip filter) to an unbalanced circuit (coaxial cable). This DC-compatible balun~\cite{Venkatesan2003InvestigationOT} contains six ports: two signal ports (a CPS port and a CPW port) and four dummy ports (CPS short, CPS open, CPW short, and CPW open). All six ports are matched to the same impedance (in our case, $50\Omega$) and electrical length $\lambda$. The signal at the CPW port is evenly distributed among the four dummy ports. Only the differential mode can propagate through the balun, while common-mode signals are rejected due to the opposite reflection phases of the port pairs~\cite{balun_IEEE}. The high-frequency cutoff of the balun is limited by its $\lambda/8$ mode. The open-circuit approximation also becomes invalid at higher frequencies. 

The flux filter works effectively when the input impedance is constant across all $\omega_{rf}$. When coaxial cables transmit flux signals through the flux line, the double-Y balun helps suppress filter ripples caused by impedance mismatch on the package side. Supplementary Figure~\ref{fig:filter_S21}(a) and (b) show HFSS simulations for two cases: a perfect on-chip $50\,\Omega$ excitation port, and the entire package. The resulting RF modulation responses, depicted in Supplementary Figure~\ref{fig:filter_S21}(c) by the black and blue curves, demonstrate that the double-Y balun effectively aligns the two curves at low frequencies. More significant ripples appear in the blue curve at frequencies ($>7$ GHz) higher than the balun range. Reducing the size of the double-Y balun can rapidly increase the cutoff frequency. In our design, the balun size ($\sim 0.5$ mm) is constrained by the need for additional wirebonds. For two-mode squeezing interactions with a frequency around $10$ GHz, directly fabricating airbridges on the balun and optimizing the design of CPS and CPW open ports will help achieve the desired frequency range.

\begin{figure}[t]
    \centering
    \includegraphics{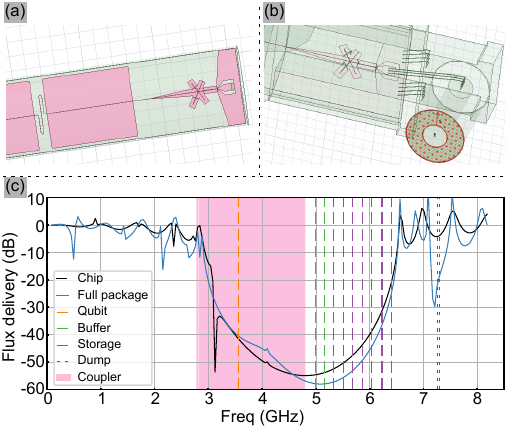}
    \caption{HFSS-simulated response of the Balun-Filter. Simulation models include (a) the coupler chip alone and (b) the complete package. (c) RF flux modulation power delivery. Black and blue curves represent different models shown in (a) and (b), which align closely for frequencies below 7 GHz. This alignment indicates compatibility between the coaxial connectors and the coupler chip. Vertical lines denote experimentally measured mode frequencies.}
    \centering
    \label{fig:filter_S21}
\end{figure}

\section{Memory cross-talk benchmarking}
\label{sec: cross-talk fitting}

Crosstalk errors can degrade the stored state fidelity during RAQM operation. These errors depend on the stage of RAQM control for a given accessed mode \(S_i\). The three stages are active access, idle, and inactive access. The dominant crosstalk errors during these stages are state-dependent access errors, many-body dephasing, and spectator access dephasing, respectively. We will present details on benchmarking these errors in the subsections below.  

\subsection{State-Dependent Access Error}
\begin{figure}[t]
    \centering
    \includegraphics{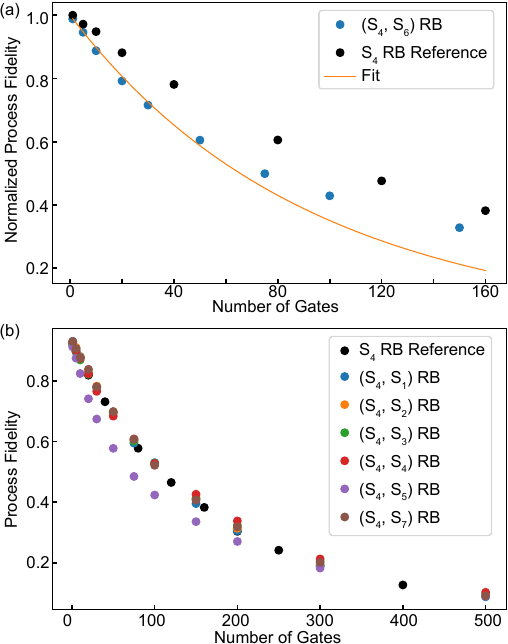}
    \caption{RB for state-dependent access error. RB is performed in the target mode $S_i$ to benchmark the $BS_i$ swap fidelity while the spectator mode $S_j$ is occupied. The decay curve for each RB (target \(S_i\), spectator \(S_j\)) is additionally averaged over $6$ cases, one for each of the possible spectator occupation: \(\{\ket{0}, \ket{1}, \ket{+x}, \ket{-x}, \ket{+y}, \ket{-y}\}\). (a) Semiclassical fit for the \((S_i, S_j) = (S_4, S_6)\) RB, reporting the additional swap error $\varepsilon_{i}^{j} = 0.938 \pm 0.060 \%$. (b) Benchmarking \(B_1S_i = B_1S_4\) swap fidelity in the presence of random population in \(S_j\).  The reference data refers to no population in the spectator mode.}
    \centering
    \label{fig:state_dep_access}
\end{figure}

The state-dependent access error occurs when fetching the target mode via the \(BS_i\) swap. When the other storage modes $S_j, (j\neq i)$ are occupied, the buffer-storage cross-Kerr interactions \(\chi_{BS_j}\) can change the \(BS_i\) swap frequency randomly. Those randomly applied off-resonant swaps have lower fidelity and finally dephase the state in $S_i$ after rounds of read/write. To benchmark this error, we perform randomized benchmarking (RB) introduced in the main text Section~III between $B_1$ and $S_i$ in the presence of populations in distinct spectator mode $S_j$. During the RAM RB experiments, the RAQM is populated with different Clifford states in the \(\{\ket{0}, \ket{1}\}\) subspace. We thus perform $6$ rounds of RB to check the $BS_i$ swap fidelity with the $6$ distinct Clifford states in $S_j$. The final RB decay curve is average over the $6$ curves. We evaluated fidelities for $15$ different gate depths in the range \([0,500]\). For each depth, we chose $20$ different sequences for each RB gate depth, with each sequence measured $1000 -5000$ times depending on gate depth (more repetitions for larger gate depths).

To extract the state-dependent access error, we perform a semi-classical Markovian analysis based on the measured RB decay curve. This analysis captures the effect of decaying spectator populations, which causes the state-dependent access error to be lower for large gate depths. As shown in Figure \ref{fig:state_dep_access} (b), after \(300\) gates (\(\sim450 \mu\)s), decay curves for all target-spectator \(S_iS_j\) pairs follow the same trend as the reference case. We will now outline the analysis procedure. 

First, the fidelity of $ B_1 \leftrightarrow S_i $ swap is state-dependent: when $ \ket{S_j} = \ket{0} $ ($ \ket{1} $), the fidelity is $ F_0 $ ($ F_1 $). At gate depth $ N $, we label the population of $ \ket{S_j} $ in the $ \ket{0} $ ($ \ket{1} $) state as $ P_0(N) $ ($ P_1(N) $). We label the initial population of $ \ket{S_j} $ in $ \ket{1} $ as $ \zeta $. After applying each gate, the probability of $\ket{S_j}$ staying at $\ket{1}$ is $A$. Ignoring photon excitation error, $\ket{S_j}$'s population distribution at gate depth $N$ is:

\begin{align}
\left\{ \begin{array}{l}
    P_{0}(N) = 1 - A^N\zeta \\
    P_{1}(N) = A^N\zeta 
\end{array} \right.
\end{align}

Here $A = \exp(-\frac{t_{B_1 \leftrightarrow S_i}}{\tilde{T}_1^{j}})$ is calculated through two experimentally measurable quantities: average single dual-rail gate length $t_{B_1 \leftrightarrow S_i}$, and $S_j$'s $T_1$ in the presence of $B_1 \leftrightarrow S_i$ dual-rail operation $\tilde{T}_1^{j}$. In our analysis, however, we will approximate $\tilde{T}_1^{j} \approx T_1^j$ as \(S_j\)'s raw \(T_1\).The measured RB process fidelity $\tilde{F}(N)$ at each gate depth $N$ is:

\begin{align}
\tilde{F}(N) &= \prod_{j=1}^{N}\left(F_0P_0(j)+F_1P_1(j)\right) \nonumber \\
&= F_0^N\prod_{j=1}^{N}\left(1+(\frac{F_1}{F_0}-1)A^j\zeta\right)
\end{align}

We define $\varepsilon_{i}^{j} \equiv 1-\frac{F_1}{F_0}$, whose absolute value is relative deviation of \(F_1\) from \(F_0\). Here, indices \(i,j\) denote the target and spectator mode, respectively. We expect \(\varepsilon_{i}^{j}\) to be a small quantity because the buffer-storage cross-Kerr interaction is a small quantity on the timescale of gates, i.e., \(\chi_{BS_j}t_{B_1 \leftrightarrow S_i}<0.5\%\). Thus, we can make the approximation $1-\varepsilon_{i}^{j} A^j\zeta\simeq \exp(-\varepsilon_{i}^{j} A^j\zeta)$ for small gate depths (\(jt_{B_1 \leftrightarrow S_i}\ll T_1^j\)). Using this approximation, we have:

\begin{align}
\tilde{F}(N) &\simeq F_0^N\exp(\varepsilon_{i}^{j}\zeta\sum_{j=1}^{N}A^j) \nonumber \\
&= F_0^N \exp\left(\varepsilon_{i}^{j}\zeta \frac{A(1-A^N )}{1-A}\right)
\label{app:cross_talk_fidelity}
\end{align}

To extract \(\varepsilon_{i}^{j}\), we use the above equation to fit the previously obtained decay curve to a depth up to \(50\) gates (\(\sim 50-75 \mu\mathrm{s}\)). We set \(\zeta = 1/2\) as the decay curve averages over the spectator mode's population over the $6$ cardinal states of its Bloch sphere. An example of this fitting is shown in Supplementary Figure~\ref{fig:state_dep_access} (a). The state-dependent access error for a swap gate is \(\varepsilon_{ai}^{j} = (1-F_0^i\varepsilon_{i}^{j})^{3/2}\), which is plotted in the main text Fig.~4 (c). For most target-spectator mode pairs, the state-dependent access error is below \(<0.3\%\), while only a few pairs have an error rate as high as \(1.396\%\). 

\subsection{Many-Body Dephasing Error} 
\label{sec: MBD stored states}

Many-body dephasing refers to $S_i$ being dephased by states in other storage modes via the storage-storage cross-Kerr interactions. This error occurs during all RAQM control stages but it is the dominant crosstalk error during the idle stage. We quantify this error by measuring the Ramsey of the target mode \(S_i\) while other storage modes are occupied. The protocol is as follows: (1) Prepare other storage modes \(\{S_j| j\neq i\}\) uniformly in one of the $6$ basis states \(\{\ket{0}, \ket{1}, \ket{+x}, \ket{-x}, \ket{+y}, \ket{-y}\}\) (2) Perform cavity Ramsey on the storage mode \(S_i\). By fitting the Ramsey signal to exponentially decaying sinusoids, we attain $6$ dephasing rates \(\{\kappa_p\}\), one for each basis state. Since the cross-Kerr interactions between storage modes ($\approx50$ Hz) are $2$ orders of magnitude smaller than the modes' decoherence rates, the \(S_i\)'s many-body dephasing rate \(\Gamma_{mbd}^{i}\) is much smaller than the modes' intrinsic dephasing rates. As the worst case, we take the many-body dephasing rate as the largest variation in dephasing rates measured among $6$ cases: 
\begin{align}
\Gamma_{mbd}^{i} &= \text{max}\{\kappa_p\} - \text{min}\{\kappa_p\}.
\end{align}
We measured this rate to be at most \(\Gamma_{mbd}^{i} = 0.510 \pm 0.222\,\mathrm{kHz}\), which introduces at most \(0.102\%\) storage read/write error.

\subsection{Spectator Access Dephasing}
Spectator access dephasing occurs when we fetch states into the buffer mode from \(S_i's\) spectator modes via \(BS_j\) swaps. Since we introduce random populations in the buffer mode, these populations effectively dephase the stored state in \(S_i\) via the \(\chi_{BS_i}\) cross-Kerr interactions. To quantify this error, we measure the Ramsey of \(S_i\) mode in the presence of random \(B_1-S_j\) beamsplitter gates. We first initialize a distinct spectator mode \(S_j\) with a single photon. We then perform Ramsey experiment on \(S_i\) where the two \(X_{\pi/2}\) pulses on \(S_i\) are separated by an RB sequence of \(BS_j\) beamsplitter gates. If we replace the RB sequence with an even number of \(BS_j\) swap gates, we would observe beating in \(S_i\)'s Ramsey trace. This beating is due to the difference between \(\chi_{BS_i}\) and \(\chi_{S_iS_j}\) cross-Kerr interactions. However, the RB sequence of \(BS_j\) beamsplitter gates randomizes the population in buffer mode, rendering the beating incoherent. To extract the spectator access dephasing rate, we subtract \(S_i\)'s intrinsic dephasing rate \(\Gamma_{\phi, \text{T2}}\) from the measured rate \(\Gamma_{\text{meas}}\). The fitted error rate \(\Gamma_{\text{meas}}\) yields the effective error rate due to \(B_1-S_j\) gates as 
\[\Gamma_{ai}^{j} = \Gamma_{\text{meas}} - \Gamma_{\phi, \text{T2}}\]
As shown in the main text Fig.~4 (d), the spectator access dephasing rate is below \(3\,\mathrm{kHz}\) for most target-spectator pairs, while some have as high as \(18.75\,\mathrm{kHz}\). As RAQM size increases, each mode spends more time idling per memory flashing round. As a result, the modes become more sensitive to spectator access dephasing as we scale the size of RAQM.

\section{Single Mode Control}
\label{sec:single_mode_control}
Using the transmon-buffer 4-wave mixing interactions, we can perform universal gates on the buffer mode in the $\{\ket{0}, \ket{1}\}$ subspace. Each buffer mode gate is a cascaded gate composed of three periods: swapping the buffer state into the transmon, performing a single transmon gate, and swapping the transmon state back into the buffer. 

The transmon-buffer swap is realized using the $\ket{f0}\leftrightarrow\ket{g1}$ sideband. The sideband is activated by charge driving with a frequency $\omega_{d4}$. Here, $\omega_{d4}\approx 2\omega_{q}+\alpha-\omega_{bi}$ is chosen based on an experimental scan of the resonant swapping frequency. We use a Gaussian flat-top pulse with a $3\sigma (\sigma=0.005\,\mu\mathrm{s})$ ramping as the pulse waveform. 

\begin{figure}[t]
    \centering
    \includegraphics{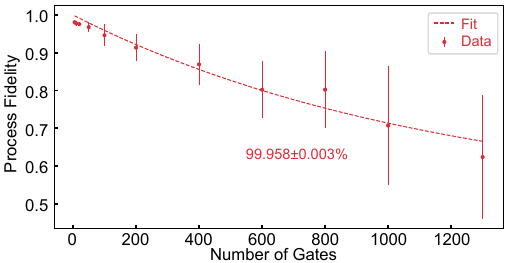}
    \caption{Transmon single qubit gate RB. The gate depths are limited by the RFSoC register memory.}
    \centering
    \label{fig:qubit_rb}
\end{figure}

After the swap, the buffer state $x\ket{0}+y\ket{1}$ is mapped to the transmon state $x\ket{g}+ye^{1j\phi_{d4}}\ket{1}$. Performing a single transmon gate requires a virtual phase correction to reverse the deterministic phase $\phi_{d4}$ coming from the sideband stark-shift. The phase cancellation procedure is similar to that described in Supplementary Section~\ref{sec:phase calibration}. After performing the single transmon gate, the state is swapped back to the buffer. Supplementary Figure~\ref{fig:qubit_rb} shows the single transmon gate fidelity ($99.958\%\pm0.003\%$) measured through RB.

The single buffer gate length is $1.65\,\mu\mathrm{s}$ with an average Clifford gate fidelity of $98.64\%$. This includes two $0.615\,\mu\mathrm{s}$ buffer-transmon swaps, two $140\,\mathrm{ns}$ transmon $\ket{f}\leftrightarrow\ket{e}$ $\pi$ pulses, and a $140\,\mathrm{ns}$ transmon $\ket{e}\leftrightarrow\ket{g}$ pulse. Compared to the small transmon-buffer $\chi$-shift ($1/\chi_{qb}\sim3.5\,\mu\mathrm{s}$), this is a cavity gate scheme beyond the $\chi$-limit. Our experimental gate speed is limited by the room-temperature amplifier power on the drive lines. 

\section{Random Read Fidelity}\label{app: random_read_fidelity}
The Random read fidelity $\bar{F}_{BS_i}$ is measured through multiplexed RB across all memory modes. For each mode $S_i$, the depolarization parameter $P_{S_i}$ is extracted by fitting the process fidelity after different RAQM cycles. For a given mode \(S_i\), a single cycle in a size-7 memory consists of 14 gates: two gates (read and write) are performed on $S_i$, while the other 12  gates are performed on the other storage modes. The latter are effectively identity gates on $S_i$. However, due to decoherence and modes' cross-Kerrs, these identity gates have nonzero infidelities. $S_i$'s random read fidelity $\bar{F}_{BS_i}$ is  
\begin{align}
\bar{F}_{BS_i}=\sqrt{1-\frac{2}{3}\left(1-\left(\frac{P_{S_i}}{P_B}\right)^{1/14}\right)}\nonumber
\end{align}
Here, we isolate the buffer mode's control fidelity, with its depolarization parameter defined as $P_B$. $P_B$ is measured independently using single-mode RB on the buffer, involving only $G_i^n$. The factor of $2/3$ arises because three levels are used in the RAQM controlling transmon~\cite{rb_knill}, and the transmon $\pi_{ef}$ rotation spreads error population from $\ket{e}$ to $\ket{f}$. This factor thus provides a conservative estimate of the fidelity.

\section{Error analysis}
\label{app:errorbudget}
\subsection{Single-mode error budget}

\begin{figure}[t]
    \centering
    \includegraphics{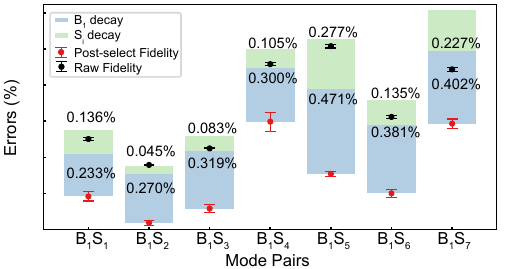}
    \caption{Error budget for $B_1S_i$ swap fidelity. The blue and green parts represent the $T_1$ decay error from the $B_1$ and $S_i$ during the swap.}
    \centering
    \label{fig:bare_mode_budget}
\end{figure}

In our buffer-storage Randomized Benchmarking experiment, each experimental point corresponds to the average of 30 random sequences, and each sequence is repeated from $1000$ to $20000$ times to get enough statistics. Each RB gate uses a Gaussian flat-top pulse with $3\sigma (\sigma=0.005\,\mu\mathrm{s})$. Since we use parity measurement to extract the buffer-storage state after finishing an RB sequence, it is possible to misassign a higher fock state population in the readout. Noticing that the buffer populations in $\ket{2}$ are treated as $\ket{0}$, the measured raw RB fidelity is still accurate if we assume negligible buffer populations beyond $\ket{3}$. Based on our low system temperature (See Supplementary Section~\ref{app:temp}), the measurement misassignment error is not a major source of error in our system.

Supplementary Figure~\ref{fig:bare_mode_budget} shows the error budget for the swap fidelity of all $B_1S_i$ pairs. The raw and post-select fidelities are all experimentally measured through randomized benchmarking. The blue and green parts represent the $T_1$ decay error from the $B_1$ and $S_i$ during the swap. We use the $B_1$ and $S_i$'s coherence during idle time (Supplementary Table~\ref{table:frequency}) to calculate the decay error contribution during beamsplitter interactions. The remaining mismatch in fidelities could come from modes' $T_1$ fluctuations. The post-selected fidelity may be limited by coupler heating and cavity dephasing.

\subsection{Full RAQM error budget}

\begin{figure}[t]
    \centering
    \includegraphics{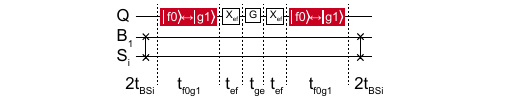}
    \caption{Pulse sequence of a storage access gate. $X_{ef}$ refers to the $\pi$ pulse on the transmon $\{\ket{e}, \ket{f}\}$ subspace. $G$ represents a random gate on the transmon $\{\ket{g}, \ket{e}\}$ subspace with the same gate length $t_{ge}$.}
    \centering
    \label{fig:rbam_error_budget}
\end{figure}

Here, we discuss the error budget for random read infidelity. Supplementary Figure~\ref{fig:rbam_error_budget} shows the pulse sequence for each $S_i$'s access gate. Here, we label the gate lengths for a $B_1S_i$ swap, a $QB_1$ $\ket{f0}\leftrightarrow\ket{g1}$ swap, 
a transmon $\ket{e}\leftrightarrow\ket{f}$ $\pi$ pulse, and a random transmon gate $G$ on the $\{\ket{g}, \ket{e}\}$ subspace as $2t_{BS_i}$, $t_{f0g1}$, $t_{ef}$, and $t_{ge}$. While $t_{BS_i}$ are dependent on \(S_i\), the other gate lengths are fixed in our experiments: $t_{f0g1}=0.606\,\mu\mathrm{s}$ and $t_{ef}=t_{ge}=0.14\,\mu\mathrm{s}$. Each $S_i$'s access gate is divided into read, transmon processing, and write periods with length $t_{\text{read}}$, $t_{\text{write}}$, and $t_{\text{proc}}$. Information is transferred from $S_i$ to $B_1$ during the read period, while the write period reverses this process. Under this definition, we have:
\begin{align}
t_{\text{read}} &= t_{\text{write}}= 2t_{BS_i} \nonumber \\
t_{\text{proc}} &= 2t_{f0g1}+2t_{ef}+t_{ge} \nonumber 
\end{align}

\begin{table*}[t]
  \begin{tabular}{c c c c c c c c}
		    \hline
		    \hline
		    Occupied modes & $S_1$ & $S_2$ & $S_3$ & $S_4$ & $S_5$ & $S_6$ & $S_7$ \\ \hline\hline 
            Active access $BS_1$ &  & $\begin{array}{c}
            0.697\% \\ \pm 0.176\%
            \end{array}$ & $\begin{array}{c}
            0.178\% \\ \pm 0.039\%
            \end{array}$ & $\begin{array}{c}
            0.019\% \\ \pm 0.031\%
            \end{array}$ & $\begin{array}{c}
            0.162\% \\ \pm 0.02\%
            \end{array}$ & $\begin{array}{c}
            0.017\% \\ \pm 0.032\%
            \end{array}$ & $\begin{array}{c}
            0.068\% \\ \pm 0.019\%
            \end{array}$ \\ \hline
            Active access $BS_2$ & $\begin{array}{c}
            0.031\% \\ \pm 0.018\%
            \end{array}$ &  & $\begin{array}{c}
            1.111\% \\ \pm 0.248\%
            \end{array}$ & $\begin{array}{c}
            0.098\% \\ \pm 0.029\%
            \end{array}$ & $\begin{array}{c}
            0.049\% \\ \pm 0.023\%
            \end{array}$ & $\begin{array}{c}
            0.236\% \\ \pm 0.063\%
            \end{array}$ & $\begin{array}{c}
            0.205\% \\ \pm 0.034\%
            \end{array}$ \\ \hline
            Active access $BS_3$ & $\begin{array}{c}
            0.108\% \\ \pm 0.028\%
            \end{array}$ & $\begin{array}{c}
            0.086\% \\ \pm 0.054\%
            \end{array}$ &  & $\begin{array}{c}
            0.055\% \\ \pm 0.027\%
            \end{array}$ & $\begin{array}{c}
            0.03\% \\ \pm 0.031\%
            \end{array}$ & $\begin{array}{c}
            0.058\% \\ \pm 0.027\%
            \end{array}$ & $\begin{array}{c}
            0.067\% \\ \pm 0.045\%
            \end{array}$ \\ \hline
            Active access $BS_4$ & $\begin{array}{c}
            0.058\% \\ \pm 0.065\%
            \end{array}$ & $\begin{array}{c}
            0.088\% \\ \pm 0.06\%
            \end{array}$ & $\begin{array}{c}
            0.04\% \\ \pm 0.012\%
            \end{array}$ &  & $\begin{array}{c}
            0.064\% \\ \pm 0.048\%
            \end{array}$ & $\begin{array}{c}
            1.396\% \\ \pm 0.089\%
            \end{array}$ & $\begin{array}{c}
            0.125\% \\ \pm 0.065\%
            \end{array}$ \\ \hline
            Active access $BS_5$ & $\begin{array}{c}
            0.179\% \\ \pm 0.071\%
            \end{array}$ & $\begin{array}{c}
            0.058\% \\ \pm 0.032\%
            \end{array}$ & $\begin{array}{c}
            0.012\% \\ \pm 0.038\%
            \end{array}$ & $\begin{array}{c}
            0.068\% \\ \pm 0.044\%
            \end{array}$ &  & $\begin{array}{c}
            0.019\% \\ \pm 0.025\%
            \end{array}$ & $\begin{array}{c}
            0.056\% \\ \pm 0.049\%
            \end{array}$ \\ \hline
            Active access $BS_6$ & $\begin{array}{c}
            0.009\% \\ \pm 0.157\%
            \end{array}$ & $\begin{array}{c}
            0.257\% \\ \pm 0.107\%
            \end{array}$ & $\begin{array}{c}
            0.091\% \\ \pm 0.061\%
            \end{array}$ & $\begin{array}{c}
            0.321\% \\ \pm 0.122\%
            \end{array}$ & $\begin{array}{c}
            0.361\% \\ \pm 0.138\%
            \end{array}$ &  & $\begin{array}{c}
            0.122\% \\ \pm 0.045\%
            \end{array}$ \\ \hline
            Active access $BS_7$ & $\begin{array}{c}
            0.011\% \\ \pm 0.052\%
            \end{array}$ & $\begin{array}{c}
            0.618\% \\ \pm 0.018\%
            \end{array}$ & $\begin{array}{c}
            0.083\% \\ \pm 0.04\%
            \end{array}$ & $\begin{array}{c}
            0.05\% \\ \pm 0.026\%
            \end{array}$ & $\begin{array}{c}
            0.11\% \\ \pm 0.049\%
            \end{array}$ & $\begin{array}{c}
            0.014\% \\ \pm 0.025\%
            \end{array}$ &  \\ 
            \hline\hline
		\end{tabular}
		\caption{RAQM state-dependent access error for storage-buffer swap.}
		\label{table:state-dependent_error}
\end{table*}

The state-dependent access error, which happens during actively accessing $S_i$, is calculated using the measured worst-case swap infidelity $\varepsilon_{ai}^{j}$:
\begin{align}
1-\left(1-\sum_{i=1;i\neq j}^{7}\varepsilon_{ai}^{j}\right)^{1/7}.
\end{align}
The coefficient accounts for the number of storage modes ($7$) in the multiplexed RAQM operation to avoid double counting (the same for the rest of the error budget discussions). Supplementary Table~\ref{table:state-dependent_error} shows the experimentally measured state-dependent access error.

Many-body dephasing occurs whenever $S_i$ is populated with the target state. Using the many-body dephasing rate $\Gamma_{mbd}^{i}$ extracted during the RAQM idle period, we can calculate the many-body dephasing error for $S_i$:
\begin{align}
1-\exp{\left(-\frac{\Gamma_{mbd}^{i}}{7}\sum_{j=1;j\neq i}^{t}\left(t_{\text{read}}+t_{\text{proc}}/2\right)\right)}.
\end{align}

\begin{table*}[t]
  \begin{tabular}{c c c c c c c c}
		    \hline
		    \hline
		    Dephasing rate (kHz) & $S_1$ & $S_2$ & $S_3$ & $S_4$ & $S_5$ & $S_6$ & $S_7$ \\ \hline\hline 
            Inactive access $BS_1$ &  & $\begin{array}{c}
            6.304 \\ \pm 0.007
            \end{array}$ & $\begin{array}{c}
            2.061 \\ \pm 0.012
            \end{array}$ & $\begin{array}{c}
            4.343 \\ \pm 0.007
            \end{array}$ & $\begin{array}{c}
            0.169 \\ \pm 0.036
            \end{array}$ & $\begin{array}{c}
            0.906 \\ \pm 0.032
            \end{array}$ & $\begin{array}{c}
            0.993 \\ \pm 0.021
            \end{array}$ \\ \hline
            Inactive access $BS_2$ & $\begin{array}{c}
            0.237 \\ \pm 0.005
            \end{array}$ &  & $\begin{array}{c}
            6.998 \\ \pm 0.018
            \end{array}$ & $\begin{array}{c}
            0.988 \\ \pm 0.017
            \end{array}$ & $\begin{array}{c}
            0.165 \\ \pm 0.023
            \end{array}$ & $\begin{array}{c}
            1.280 \\ \pm 0.030
            \end{array}$ & $\begin{array}{c}
            0.854 \\ \pm 0.019
            \end{array}$ \\ \hline
            Inactive access $BS_3$ & $\begin{array}{c}
            1.089 \\ \pm 0.003
            \end{array}$ & $\begin{array}{c}
            0.226 \\ \pm 0.009
            \end{array}$ &  & $\begin{array}{c}
            1.363 \\ \pm 0.027
            \end{array}$ & $\begin{array}{c}
            0.256 \\ \pm 0.021
            \end{array}$ & $\begin{array}{c}
            1.024 \\ \pm 0.033
            \end{array}$ & $\begin{array}{c}
            0.510\\ \pm 0.026
            \end{array}$ \\ \hline
            Inactive access $BS_4$ & $\begin{array}{c}
            7.182 \\ \pm 0.002
            \end{array}$ & $\begin{array}{c}
            2.464 \\ \pm 0.004
            \end{array}$ & $\begin{array}{c}
            2.288 \\ \pm 0.006
            \end{array}$ &  & $\begin{array}{c}
            2.940 \\ \pm 0.005
            \end{array}$ & $\begin{array}{c}
            18.75 \\ \pm 0.033
            \end{array}$ & $\begin{array}{c}
            2.763 \\ \pm 0.007
            \end{array}$ \\ \hline
            Inactive access $BS_5$ & $\begin{array}{c}
            2.625 \\ \pm 0.002
            \end{array}$ & $\begin{array}{c}
            0.195 \\ \pm 0.007
            \end{array}$ & $\begin{array}{c}
            1.698 \\ \pm 0.027
            \end{array}$ & $\begin{array}{c}
            0.987 \\ \pm 0.011
            \end{array}$ &  & $\begin{array}{c}
            0.842 \\ \pm 0.023
            \end{array}$ & $\begin{array}{c}
            0.881 \\ \pm 0.012
            \end{array}$ \\ \hline
            Inactive access $BS_6$ & $\begin{array}{c}
            11.25 \\ \pm 0.002
            \end{array}$ & $\begin{array}{c}
            0.067 \\ \pm 0.009
            \end{array}$ & $\begin{array}{c}
            0.190 \\ \pm 0.049
            \end{array}$ & $\begin{array}{c}
            0.282 \\ \pm 0.023
            \end{array}$ & $\begin{array}{c}
            0.243 \\ \pm 0.016
            \end{array}$ &  & $\begin{array}{c}
            0.502 \\ \pm 0.019
            \end{array}$ \\ \hline
            Inactive access $BS_7$ & $\begin{array}{c}
            6.496 \\ \pm 0.002
            \end{array}$ & $\begin{array}{c}
            11.85 \\ \pm 0.001
            \end{array}$ & $\begin{array}{c}
            2.694 \\ \pm 0.017
            \end{array}$ & $\begin{array}{c}
            3.088 \\ \pm 0.004
            \end{array}$ & $\begin{array}{c}
            3.552 \\ \pm 0.004
            \end{array}$ & $\begin{array}{c}
            2.848 \\ \pm 0.009
            \end{array}$ &  \\ 
            \hline\hline
		\end{tabular}
		\caption{RAQM spectator-access dephasing rate.}
		\label{table:spectator-access_dephasing}
\end{table*}

The spectator-access dephasing, which happens during the $S_i$'s inactive-access period, is calculated using the measured $S_j$'s spectator-access dephasing rate $\Gamma_{ai}^{j}$:
\begin{align}
1-\prod_{j=1;j\neq i}^{7}\left(1-\Gamma_{ai}^{j}(t_{\text{read}}+t_{f0g1})/7\right).
\end{align}
Details about the spectator-access dephasing rate are shown in Supplementary Table~\ref{table:spectator-access_dephasing}. For each pair's RB, we use $30$ different sequences for each RB gate depth to get statistics.

The RAQM swap errors are scaled by the average number of swaps on $S_i$'s random read operations during RAM experiments:
\begin{align}
\left(1-F_{ri}^{1/7}\right).
\end{align}

\begin{table*}[t]
  \begin{tabular}{c c c c c c c c}
		    \hline
		    \hline
		    Storage mode & $S_1$ & $S_2$ & $S_3$ & $S_4$ & $S_5$ & $S_6$ & $S_7$ \\ \hline\hline 
            State-dependent access error & $0.164\%$ & $0.245\%$ & $0.058\%$ & $0.255\%$ & $0.056\%$ & $0.167\%$ & $0.127\%$ \\ \hline
            Many-body dephasing & $0.102\%$ & $0.046\%$ & $0.030\%$ & $0.026\%$ & $0.030\%$ & $0.028\%$ & $0.042\%$ \\ \hline
            Spectator-access dephasing & $0.928\%$ & $0.705\%$ & $0.450\%$ & $0.324\%$ & $0.242\%$ & $0.734\%$ & $0.177\%$ \\ \hline
            Swaps & $0.072\%$ & $0.051\%$ & $0.064\%$ & $0.132\%$ & $0.146\%$ & $0.089\%$ & $0.127\%$ \\ \hline
            Storage decay & $0.559\%$ & $0.159\%$ & $0.245\%$ & $0.329\%$ & $0.525\%$ & $0.327\%$ & $0.493\%$ \\ \hline
            Storage dephasing & $0.084\%$ & $0.054\%$ & $0.027\%$ & $0.038\%$ & $0.019\%$ & $0.020\%$ & $0.018\%$ \\
             \hline\hline
             Calculated random read infidelity &
            $\begin{array}{c}
            1.854\%
            \end{array}$ &
            $\begin{array}{c}
            1.181\%
            \end{array}$ &
            $\begin{array}{c}
            0.855\%
            \end{array}$ &
            $\begin{array}{c}
            1.019\%
            \end{array}$ &
            $\begin{array}{c}
            0.999\%
            \end{array}$ &
            $\begin{array}{c}
            1.309\%
            \end{array}$ &
            $\begin{array}{c}
            0.942\%
            \end{array}$ \\ \hline
            Measured random read infidelity &
            $\begin{array}{c}
            1.171\% \\ \pm 0.041\%
            \end{array}$ &
            $\begin{array}{c}
            1.259\% \\ \pm 0.041\%
            \end{array}$ &
            $\begin{array}{c}
            1.112\% \\ \pm 0.042\%
            \end{array}$ &
            $\begin{array}{c}
            1.114\% \\ \pm 0.049\%
            \end{array}$ &
            $\begin{array}{c}
            1.232\% \\ \pm 0.066\%
            \end{array}$ &
            $\begin{array}{c}
            1.232\% \\ \pm 0.060\%
            \end{array}$ &
            $\begin{array}{c}
            1.192\% \\ \pm 0.056\%
            \end{array}$ \\
            \hline\hline
		\end{tabular}
		\caption{RAQM random read error budget.}
		\label{table:rbam_budget}
\end{table*}

The decay and dephasing error of $S_i$ includes all the inactive-access periods for $S_i$:
\begin{align}
1-\exp{\left(-\frac{\Gamma_{1}^{i}}{7}\sum_{j=1;j\neq i}^{t}\left(t_{\text{read}}+t_{\text{proc}}/2\right)\right)}, \\
1-\exp{\left(-\frac{\Gamma_{\phi}^{i}}{7}\sum_{j=1;j\neq i}^{t}\left(t_{\text{read}}+t_{\text{proc}}/2\right)\right)}. 
\end{align}
$\Gamma_{1}^{i}$ and $\Gamma_{\phi}^{i}$ represent the decay and pure dephasing rate for $S_i$. Supplementary Table~\ref{table:rbam_budget} lists the error budget in detail. The average random read fidelities of different RAQM sizes are shown in Supplementary Table~\ref{table:all_rbam}.

For the RAM RB experiments, we chose $30$ different sequences for each RB gate depth, with each sequence measured $1000$ times to get statistics. Since transmon's $\{\ket{g}, \ket{e}, \ket{f}\}$ are in the RAM RB, the process fidelity should stabilize at $1/3$, given an evenly distributed readout contrast for all states. In our experiments, the readout signal difference in $\{\ket{g}, \ket{e}, \ket{f}\}$ causes the final process fidelity to be deviated from $1/3$.

\begin{table}[t]
  \begin{tabular}{c c}
  \hline\hline
  RAQM size & Average random read infidelity \\
  \hline\hline
  $[S_1]$ & $0.192\%\pm0.046\%$ \\ \hline   
  $[S_2]$ & $0.161\%\pm0.037\%$\\ \hline 
  $[S_3]$ & $0.299\%\pm0.043\%$\\ \hline 
  $[S_4]$ & $0.160\%\pm0.035\%$\\ \hline 
  $[S_5]$ & $0.473\%\pm0.045\%$\\ \hline 
  $[S_6]$ & $0.906\%\pm0.051\%$\\ \hline 
  $[S_7]$ & $0.590\%\pm0.048\%$\\ \hline \hline
  $[S_1, S_2]$ & $0.630\%\pm0.025\%$\\ \hline 
  $[S_1, S_3]$ & $0.612\%\pm0.021\%$\\ \hline 
  $[S_1, S_4]$ & $0.631\%\pm0.023\%$\\ \hline 
  $[S_1, S_5]$ & $1.130\%\pm0.028\%$\\ \hline 
  $[S_1, S_6]$ & $1.027\%\pm0.025\%$\\ \hline 
  $[S_1, S_7]$ & $0.588\%\pm0.018\%$\\ \hline 
  $[S_2, S_3]$ & $0.867\%\pm0.021\%$\\ \hline 
  $[S_2, S_4]$ & $0.545\%\pm0.020\%$\\ \hline 
  $[S_2, S_5]$ & $0.843\%\pm0.025\%$\\ \hline 
  $[S_2, S_6]$ & $0.975\%\pm0.022\%$\\ \hline 
  $[S_2, S_7]$ & $0.706\%\pm0.019\%$\\ \hline 
  $[S_3, S_4]$ & $0.619\%\pm0.019\%$\\ \hline 
  $[S_3, S_5]$ & $0.936\%\pm0.028\%$\\ \hline 
  $[S_3, S_6]$ & $0.936\%\pm0.028\%$\\ \hline 
  $[S_3, S_7]$ & $0.615\%\pm0.049\%$\\ \hline 
  $[S_4, S_5]$ & $0.572\%\pm0.019\%$\\ \hline 
  $[S_4, S_6]$ & $1.474\%\pm0.040\%$\\ \hline 
  $[S_4, S_7]$ & $0.755\%\pm0.021\%$\\ \hline 
  $[S_5, S_6]$ & $0.826\%\pm0.020\%$\\ \hline 
  $[S_5, S_7]$ & $1.143\%\pm0.033\%$\\ \hline 
  $[S_6, S_7]$ & $1.254\%\pm0.027\%$\\ \hline\hline 
  $[S_4, S_5, S_7]$ & $0.783\%\pm0.013\%$\\ \hline   
  $[S_1, S_2, S_3]$ & $1.016\%\pm0.017\%$\\ \hline 
  $[S_1, S_2, S_6]$ & $0.993\%\pm0.114\%$\\ \hline\hline 
  $[S_1, S_2, S_3, S_4, S_5, S_6, S_7]$ & $1.187\%\pm0.019\%$ \\ \hline\hline 
  \end{tabular}
  \caption{Average random read fidelity of different RAQM sizes.}
    \label{table:all_rbam}
\end{table}

\subsection{Master equation simulation} \label{app:simulation}
In this section, we perform a master-equation simulation to study the impact of decoherence on the fidelity of \(BS_i\) swap gates using QuTiP ~\cite{Johansson_2012_qutip}. We focused on \(BS_2\) gate, which had the lowest infidelity as shown in the main text Fig.~2 (e). We considered the following lab-frame Hamiltonian:
    $$H_{0}=\omega_{b}a_{b}^{\dagger}a_{b} +\omega_s a_s^\dagger a_s + H_c$$
where \(H_c\) contains interactions of the modes with the coupler
\begin{align}
    H_c &= \omega_{c0} \sqrt{|\cos(\varphi_{\text{ext}}(t))|} a_{c}^\dagger a_c \nonumber \\
    &+ g_{cb} (a_c^\dagger a_{b} + h.c.) + g_{cs}(a_c^\dagger a_{si} + h.c.).  
\end{align}
Here \(\omega_{c0}\) is the bare frequency of the SQUID coupler (at zero flux bias) and \(g_{ci}\) is the Jaynes-Cummings coupling between the SQUID coupler and the mode \(i\). All parameters were extracted from experiments. Here, the flux through the SQUID loop is modulated as
\begin{align}
\varphi_{\text{ext}}(t) = \frac{\pi \Phi_{\text{ext}}}{\Phi_0} = \varphi_{\text{DC}} + \epsilon\cos{(\omega t + \phi(t) )}
\end{align}
using a RF drive with strength \(\epsilon\) and frequency \(\omega\). The DC offset \(\varphi_{\text{DC}} = 0.269\pi\) is also the same as that used in experiments. We diagonalized \(H_0\) in absence of the drive (\(\epsilon = 0\)) to obtain the dressed states in the dual rail subspace of two modes \(|0\rangle_L = {|\widetilde{100}\rangle},|1\rangle_L ={|\widetilde{001}\rangle} \). Here, the ket symbol follows the order \(|\omega_b, \omega_c,\omega_{s}\rangle\). However, for numerical stability, we performed simulations in the rotating frame of the two modes and the coupler \(U = \exp{\{-i[\omega_b b^\dagger b + \omega_s s^\dagger s + \omega_c c^\dagger c)t]\}}\) where \(\omega_c = \omega_{c0} \sqrt{|\cos(\varphi_{\text{ext}}(0))|}\) is the frequency of the coupler in the absence of RF modulation. The Hamiltonian is as follows: 
\begin{align}
    H_{\text{rot}} &= UH_0U^\dagger  \nonumber \\
    &=[\omega_{c0}  \sqrt{|\cos(\varphi_{\text{ext}}(t))|} - \omega_c]a_{c}^\dagger a_c\nonumber \\
    &+ g_{cb} (e^{i\Delta_{cb}t}a_c^\dagger a_{b} + h.c.) + g_{cs}(e^{i\Delta_{cs}t}a_c^\dagger a_{si} + h.c.)    
\end{align}
where the detunings are \(\Delta_{ci} = \omega_c - \omega_i \). 

To calibrate the beamsplitter gate \(U_{\text{b.s}}(0) = X/2 \), we first selected the drive amplitude and frequency which maximized \(\mathcal{F} = |\langle0_L|1_L\rangle|^2\) after a \(X/2+ X/2\) pulse sequence. Each \(X/2 \) is a flat top pulse with flat length \(0.532 \mathrm{\,\mu s}\) and a gaussian ramp of duration \(15 \text{ ns}\) and \(\sigma = 5 \text{ ns}\), which are the same parameters as that used in the experiment. Since the drive amplitude and frequency have a nonlinear relationship, we swept over both parameters to arrive at the optimal point (\(\epsilon = 0.03505\), \(\omega = 514.1731\,\mathrm{MHz}\)) which yielded \(1-\mathcal{F} <10^{-5}\). Moreover, we added a global phase to convert the \(\sqrt{\text{iSWAP}}\) into \(\sqrt{\text{SWAP}}\) gate.

Finally, we performed a simulation of beam splitter Randomized Benchmarking~\cite{luSchoelkopf2023} in the presence of decoherence. The density matrix of system evolves according to the Lindblad master equation as \begin{align}
    \partial_t \rho = -i[H_{\text{rot}}, \rho] + \sum_{i = b,c,s}\Gamma_{1,i}\mathcal{D}[a_i]\rho + \sum_{i = b,s}\Gamma_{2,i}\mathcal{D}[a_i^\dagger a_i]\rho 
\end{align}
where \(\mathcal{D}[L] = L\rho L^\dagger -(1/2)\{L^\dagger L, \rho\} \) is the Liouvillian superoperator,  \(\Gamma_{1,i} = 1/T_{1,i}\) is the decay rate and (\(\Gamma_{2,i} = 1/T_{\text{echo},i} - 1/2T_{1,i} \)) is the pure dephasing rate. The decay and dephasing times were extracted from experiments. We did not include coupler dephasing as we weren't able to perform the Ramsey echo experiment on the coupler at \(\phi_{\text{DC}}\) due to large flux noise. We evaluated fidelities at 13 gate depths in the range \([1,100]\) with 5 different sequences for each gate depth. The simulation results are shown in Table \ref{table:me_sim_rb}. These results show that while the raw fidelity of \(BS_2\) swap gate is limited primarily by decay, its post-selected fidelity is limited by the pure dephasing of the two modes. Additionally, this simulation overestimates the gate infidelity by \(25\%\) due to numerical errors of the solver which make long simulations unfeasible.

\begin{table*}[t]
  \begin{tabular}{c c c}
  \hline\hline
  Errors & Raw Infidelity  & Post Selected Infidelity \\
  \hline\hline
  Dephasing $[\Gamma_{\phi, b}, \Gamma_{\phi, s}]$   & $0.1095\%\pm0.0019\%$ &  $0.0481\%\pm0.0005\%$\\ \hline   
  Decay $[\Gamma_{1, b}, \Gamma_{1, s}, \Gamma_{1, c} ]$ & $0.4007\%\pm0.0043\%$ &  $0.0039\%\pm0.0001\%$\\ \hline 
  Decay and Dephasing $[\Gamma_{1, b}, \Gamma_{1, s}, \Gamma_{1, c},\Gamma_{\phi, b}, \Gamma_{\phi, s}]$ & $0.4408\%\pm0.0052\%$ &  $0.0479\%\pm0.0006\%$\\ \hline \hline 
  Experiment & $0.35789\%\pm0.00002\%$ &  $0.036890\%\pm0.01520\%$\\ \hline \hline 
  \end{tabular}
  \caption{Master Equation Simulation of \(BS_2\) swap gate using beam-splitter Randomized Benchmarking\cite{luSchoelkopf2023}}
    \label{table:me_sim_rb}
\end{table*}

\section{Electrostatic Discharge Protection}
\label{app:ESD}

Our coupler chip uses a sapphire substrate and lacks a ground plane, making it highly susceptible to Electrostatic Discharge (ESD), significantly reducing fabrication yield.  The coupler’s dimensions ($43\times 6$ mm), the long flux line ($\sim 70$ mm), and the large pads in the flux line exacerbate ESD issues, resulting in almost $0\%$ yield of the coupler without ESD protections. The large pads can accumulate static charge during liftoff, and discharging generates a large current around the SQUID loop, damaging the coupler junctions.

To mitigate ESD issues, we employ two strategies: First, we increase the impedance of the high-impedance section, which broadens the filter stopband and reduces discharge current along the flux line. Second, we add three shorting lines around the coupler, creating a bypass circuit to protect the SQUID during fabrication. The shorting lines are carefully scratched with a clean tweezer during packaging, with the scratch size controlled to minimize the loss that could affect the coupler’s $T_1$.

\section{RAQM Advantage}\label{app: raqm_advantage}

We define the RAQM advantage as the reduction of control lines in a quantum computer without compromising the number of logical qubits and the logical error rate. Ideally, when all memory modes have infinitely long coherence and the memory read/write error rate is less than the error correction threshold, we can control an arbitrary number of logical qubits with a single set of control lines. Increasing the RAQM size here incurs no extra cost, as no errors accumulate while information stays in the RAQM. Hence RAQM advantage exists intuitively.

We now analytically demonstrate this advantage when a RAQM is attached to each data qubit of a given logical encoding, as shown in Fig. 1a in the main text.  We define $[n,k,d]$ as the number of physical qubits \(n\) (including ancilla qubits for stabilizer measurement), number of logical qubits \(k\), and code distance \(d\). We further assume that the quantum computer requires $k_0$ logical qubits with a logical error rate at most $\varepsilon_{\text{FT}}$. Additionally, we assume that each physical qubit requires $C_{q}$ control lines to perform error correction without RAQM, and $C_q+1$ control lines to perform error correction with RAQM.

Without RAQM attachments,  we assume that each logical module (set of data and ancilla qubits in the processor layer) employs the error correction code $[n_1, k_1, d_1]$. Here we need $k_0/k_1$ copies of the error correction code. The total number of control lines is then
\begin{align}\label{eq:enc_ratio_no_mem}
\left(\frac{k_0}{k_1}\right) \times n_1C_{q}.
\end{align}

With a RAQM of size $N$ attached to each data qubit,  we assume that each logical module employs a bigger error correction code $[n_2, k_2, d_2]$ to achieve the target logical error rate $\varepsilon_{\text{FT}}$. Since incorporating a memory will introduce additional errors, the code distance in this case is no smaller than in the memory-less case: \(d_2\geq d_1\). Using the RAQM, we only need $k_0/(k_2N)$ copies of the error correction code. Thus, the total number of lines to control $k_0$ logical qubits becomes 
\begin{align}\label{eq: enc_ratio_mem}
\left(\frac{k_0}{Nk_2}\right) \times n_2(C_{q} + 1). 
\end{align}
The RAQM advantage exists if Eq.~\ref{eq: enc_ratio_mem} is smaller than Eq.~\ref{eq:enc_ratio_no_mem} and the code $[n_2, k_2, d_2]$ enabling fault tolerance with RAQM exists.

We consider two classes of error-correcting codes combined with RAQM: (1) codes with a constant encoding ratio, and (2) the surface code. The first case represents the most efficient form of logical encoding (favorable to RAQM), while the second case is a currently existing code (unfavorable to RAQM). If the RAQM advantage exists in both cases, then any code with an encoding ratio lying between them should also exhibit a regime of RAQM advantage.

\subsection{Case 1:  Constant Encoding Ratio}

In this idealized case, even though incorporating memory requires a larger code distance, the encoding ratio remains the same as in the memory-less case: $k_1/n_1 = k_2/n_2$. Comparing Eq.~\ref{eq:enc_ratio_no_mem} and Eq.~\ref{eq: enc_ratio_mem}, if a RAQM with $N>2$ is introduced and the combined processor and memory read/write errors per logical cycle remain below threshold, a RAQM advantage exists. For example, good quantum low-density parity-check (LDPC) codes maintain a constant encoding ratio~\cite{LDPC_review}, which is precisely the regime where RAQM offers the greatest advantage.

This scenario can be understood practically. In an experimental setup with a fixed number of control lines—and thus a fixed number of physical qubits—using a code with constant encoding ratio yields only a fixed number of logical qubits. Changing the code distance does not alter this limit: both the total number of physical and logical qubits remain fixed by the available control lines. By contrast, incorporating memory allows the number of logical qubits to scale despite fixed control resources. Attaching a size-$N$ memory to each physical qubit enables roughly an $N$-fold increase in logical qubits, provided memory read/write errors remain below threshold.

\subsection{Case 2: Surface Code}

Let $t_{\text{log}}$ and $t_{\text{swap}}$ denote the error correction cycle time and RAQM read/write time. For simplicity, we further assume that the only error channel is decoherence, and all gates (including measurements) are implemented perfectly. We assume that the processor qubit and RAQM memory coherence time are separately $T_{q}$ and $T_{m}$. Without using RAQM, the surface code distance $d_1$ should satisfy
\begin{align}
    \left(\frac{p_q}{p^{\text{th}}_q}\right)^{\frac{d_1 + 1}{2}} \leq \varepsilon_{\text{FT}}
\end{align}
Here the physical error per correction cycle $p_q$ is approximated as $t_{\text{log}}/T_{q}$. $p^{\text{th}}_q$ is the error correction threshold for qubit errors. Since we only consider the decoherence error, the two-qubit gate error has the highest weight in the threshold~\cite{Acharya2024}, and we approximate $p^{\text{th}}_q$ to be $1\%$. To control $k_0$ logical qubits with target fault tolerance, the minimum number of control lines per logical qubit is around 
\begin{align}
2k_0d_1^2C_q.
\label{eq:woraqm}
\end{align}

When an RAQM of size $N$ is introduced, the additional physical error rate $p_m$ per correction cycle due to the memory  is:
\begin{align}
p_m\approx\frac{(N-1)(2t_{\text{swap}}+t_{\text{log}})}{T_m}+ 2t_{\text{swap}} \times \frac{1}{2}\left(\frac{1}{T_{m}}+\frac{1}{T_q}\right).
\end{align}
where the first term represents the accumulated memory idle error, and the latter two terms represent the gate error associated with memory read and write. For the gate error, we assume that the state spends equal time in the qubit and the memory mode, and thus the error includes contributions from both. Here, $p_m$ errors are analogous to qubit idle errors~\cite{Acharya2024}, which carry a different weight than two-qubit gate errors and therefore correspond to a different error-correction threshold. Here, we take $p^{\text{th}}_m = 16\%$ as the threshold for memory errors. We note that this value is decoder-dependent and idealized, but the precise number does not affect the discussion below. The minimum surface code distance $d_2$ should satisfy:
\begin{align}
\left(\frac{p_q}{p^{\text{th}}_q}+\frac{p_m}{p^{\text{th}}_m}\right)^{\frac{d_2 + 1}{2}} \leq \varepsilon_{\text{FT}} 
\end{align}
and the minimum number of control lines per logical qubit is 
\begin{align}
2k_0d_2^2(C_q+1)/N.
\label{eq:wraqm}
\end{align}
\begin{figure}[t]
    \centering
    \includegraphics{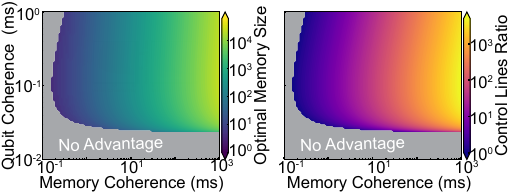}
    \caption{Approximated memory size (left) that maximizes reduction in control lines, and thus yields the highest corresponding memory-less to memory-full control line ratio (right), demonstrating RAQM advantage with the surface code. The approximation is based on the parameters $\{t_{\text{log}}, t_{\text{swap}}, {p^{\text{th}}_q}, {p^{\text{th}}_m}, C_q\} = \{200\,\text{ns}, 500\,\text{ns}, 1\%, 16\%, 3\}$ which are slightly pessimistic towards RAQM. Note that the current state-of-the-art parameters are $t_{\text{log}} \approx 1\mathrm{\mu s} $ \cite{Acharya2024} and $t_{\text{swap}}\approx 100 \mathrm{n s}$ \cite{luSchoelkopf2023}. When the RAQM size becomes sufficiently large, additional couplers may be required.}
    \centering
    \label{fig:raqm_advantage}
\end{figure}
When comparing the control lines in memory-less (Eq.~\ref{eq:woraqm}) and memory-full (Eq.~\ref{eq:wraqm}) cases, there exists both a lower and an upper bound to RAQM size for achieving an advantage with the surface code. Intuitively, if the memory contains too few modes, the higher error rate of the swap gate outweighs the benefit of increased storage coherence. Conversely, if the memory is too large, each mode spends considerable time idling while others are being processed, and correcting the resulting accumulated idling errors may require more lines than in the memory-less case. 

For given coherence times of the qubit and the memory, we plot both the optimal RAQM size $N$ and control lines ratio (memory-less/memory-full) in Supplemental Figure~\ref{fig:raqm_advantage}. As shown in the figure, achieving a RAQM advantage requires the memory's coherence time to be approximately $1$ to $2$ orders of magnitude longer than that of the qubit processor. Using state-of-the-art coherence times $T_{q}= 0.1\mathrm{ms}$ and $T_{m}= 10\mathrm{ms}$, we can use around $200$ modes in the RAQM to control the same amount of logical qubits with the same logical error rate, while needing $50$ times fewer control lines. As the RAQM size increases, additional couplers may be required, so that the ultimate scaling may vary slightly depending on the specific implementation.

Additionally, unlike classical RAM, where computations slow down due to the extra time required to transfer data between memory and processor, RAQM does not have a similar disadvantage in logical computation speed when heavy error correction code is needed. When performing logical entangling gates in surface code, lattice surgery takes $\text{O}(d_1)$ code cycles~\cite{Herr_2017}. With transversal entangling gates enabled by RAQM, the code cycle cost scales as $\text{O}(t_{\text{swap}}/t_{\text{log}})$, which is not necessarily greater than that in the memory-less case.

\section{Device Fabrication and Measurement Setup}

Our Casaded random-accessed quantum memory has three components: the multimode flute cavity, the transmon chip, and the coupler chip. 

The multimode cavity is fabricated from high-purity (99.9995\%) Aluminum block with dimensions $7.6 \,\text{cm} \times 6.985 \,\text{cm} \times 12.522 \,\text{cm}$. Three rounds of chemical etching removed in total $\sim 200 \mu$m surface using Transene Aluminum Etchant A.

For the transmon chip, we use a 430~$\mu$m thick C-plane HEMEX sapphire wafer annealed at $1200^o$C for 2 hours as the substrate. 150~nm thick Tantalum film is sputtered at $800^o$C as the ground plane. Large patterns, except Josephson junctions, are fabricated through photolithography with Heidelberg MLA 150 Direct Writer and fluorine-based dry-etching. The junction mask was fabricated with the Raith EBPG 5000+ E-Beam writer on a bi-layer resist (MMA EL11-950 PMMA A7). Transmon's Josephson junctions are of the Manhattan type. The junction mask was metalized in a Plassys electron beam evaporator with double-angle evaporation. The wafer was then diced and lifted off and packaged inside the multimode flute cavity. 

For the coupler chip, we use a 530~$\mu$m thick EFG C-plane sapphire wafer annealed at $1200^o$C for 2 hours as the substrate. 150~nm thick Niobium film is evaporated in Kurt J. Lesker E-beam Evaporator at room temperature as the ground plane. Large patterns, except Josephson junctions, are fabricated through photolithography with Heidelberg MLA 150 Direct Writer and Fluorine-based dry-etching. The junction mask was fabricated with Raith EBPG 5200+ E-Beam writer on a bi-layer resist (MMA EL13-950 PMMA A4). Coupler's Josephson junctions are Dolan bridge type. The junction mask was metalized in a Plassys electron beam evaporator with double-angle evaporation. The wafer was then diced and lifted off. Each shorting line structure mentioned in Supplementary Section ~\ref{app:ESD} is scratched six times with an IPA-cleaned tweezer during packaging.

\begin{figure*}[t]
    \centering
    \includegraphics{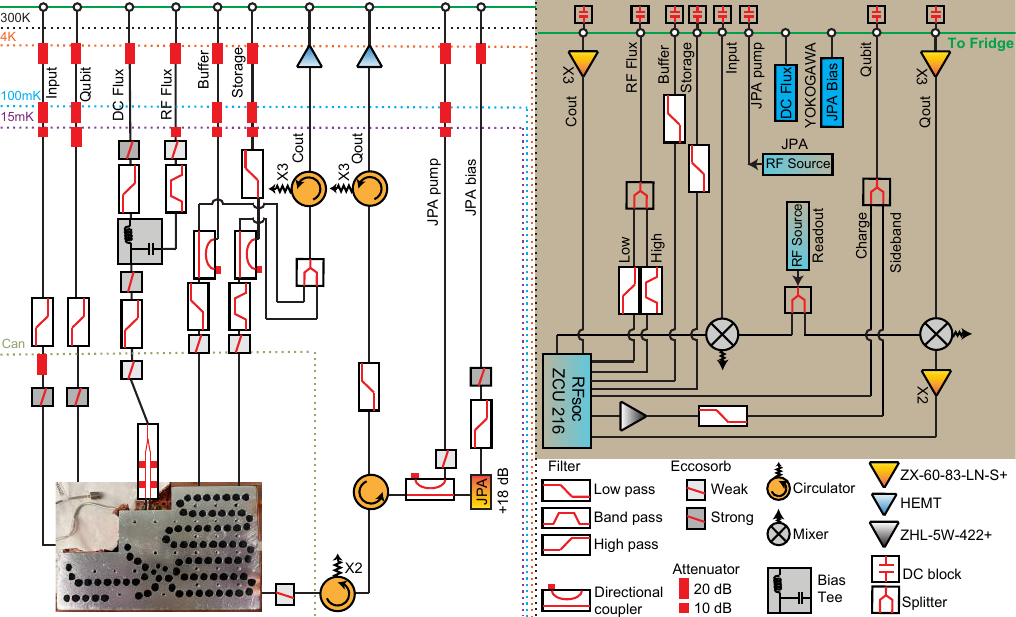}
    \caption{Detailed cryogenic and room temperature measurement setup.}
    \centering
    \label{fig:measurement}
\end{figure*}

The multimode device is packaged inside a double layer $\mu$-metal shielded sample can and installed inside a dilution fridge. 

Supplementary Figure~\ref{fig:measurement} shows the room and cryogenic temperature measurement chain. The device is mounted on the mixing chamber plate of a dilution refrigerator with a base temperature of 10 mK. All RF input signals are generated via the ZCU216 RFSoC board. Two DC sources (Yokogawa GS200) bias the DC flux of the coupler and the Josephson Parametric Amplifiers (JPA). One Signal Core SC5511A serves as the local oscillator (LO), split into two tones: one tone, after frequency up-conversion, functions as the readout pulse; the other is used for frequency down-conversion and is collected by the RFSoC ADC. A second Signal Core SC5511A provides the pump tone for the JPA, amplifying the qubit readout signal (labeled as Qout). The pulse sent to the qubit channel includes two components: the single-qubit drive and the qubit-buffer sideband drive, both synthesized directly through different RFSoC DAC channels. The sideband drive is further amplified with a ZHL-5W-422+ high-power amplifier at room temperature. Buffer and Storage modes are probed separately through a weakly coupled port machined at the multimode cavity; the signals are then combined and collected by the RFSoC ADC. Due to the RFSoC Nyquist frequency range, RF flux modulation is synthesized through two channels: low modulation frequency (0.1–1 GHz) and high modulation frequency (1–3 GHz). These two signals are combined at room temperature and then sent through the RF flux line, where they are combined with the DC flux at base temperature. The signal reaches the SMA connector, which is wire-bonded directly to the coupler chip. All RF lines have a DC block at the top of the fridge to avoid ground-loop, and all lines coming out of the device have Quantum Microwave eccosorbs placed as close as possible to the sample.

\bibliography{2Q}

@misc{zhou2025lowoverheadtransversalfaulttolerance,
      title={Low-Overhead Transversal Fault Tolerance for Universal Quantum Computation}, 
      author={Hengyun Zhou and Chen Zhao and Madelyn Cain and Dolev Bluvstein and Nishad Maskara and Casey Duckering and Hong-Ye Hu and Sheng-Tao Wang and Aleksander Kubica and Mikhail D. Lukin},
      year={2025},
      eprint={2406.17653},
      archivePrefix={arXiv},
      primaryClass={quant-ph},
      url={https://arxiv.org/abs/2406.17653}, 
}

@Article{Beccari2022,
author={Beccari, A.
and Visani, D. A.
and Fedorov, S. A.
and Bereyhi, M. J.
and Boureau, V.
and Engelsen, N. J.
and Kippenberg, T. J.},
title={Strained crystalline nanomechanical resonators with quality factors above 10 billion},
journal={Nature Physics},
year={2022},
month={Apr},
day={01},
volume={18},
number={4},
pages={436-441},
issn={1745-2481},
doi={10.1038/s41567-021-01498-4},
url={https://doi.org/10.1038/s41567-021-01498-4}
}

@article{nanomech_ultra_high,
  title = {Ultrahigh Quality Factor of a Levitated Nanomechanical Oscillator},
  author = {Dania, Lorenzo and Bykov, Dmitry S. and Goschin, Florian and Teller, Markus and Kassid, Abderrahmane and Northup, Tracy E.},
  journal = {Phys. Rev. Lett.},
  volume = {132},
  issue = {13},
  pages = {133602},
  numpages = {7},
  year = {2024},
  month = {Mar},
  publisher = {American Physical Society},
  doi = {10.1103/PhysRevLett.132.133602},
  url = {https://link.aps.org/doi/10.1103/PhysRevLett.132.133602}
}

@article{Ruskuc2022NuclearSpinWave,
  author       = {Ruskuc, Andrei and Wu, Chun-Ju and Rochman, Jake and Choi, Joonhee and Faraon, Andrei},
  title        = {Nuclear spin-wave quantum register for a solid-state qubit},
  journal      = {Nature},
  volume       = {602},
  number       = {7897},
  pages        = {408--413},
  year         = {2022},
  doi          = {10.1038/s41586-021-04293-6}
}

@article{MacCabe2020NanoAcoustic,
  author       = {MacCabe, Gregory S. and Ren, Hengjiang and Luo, Jie and Cohen, Justin D. and Zhou, Hengyun and Sipahigil, Alp and Mirhosseini, Mohammad and Painter, Oskar},
  title        = {Nano-acoustic resonator with ultralong phonon lifetime},
  journal      = {Science},
  volume       = {370},
  number       = {6518},
  pages        = {840--843},
  year         = {2020},
  doi          = {10.1126/science.abc7312}
}

@article{QRAM_3d_cav_weiss,
  title = {Quantum Random Access Memory Architectures Using 3D Superconducting Cavities},
  author = {Weiss, D.K. and Puri, Shruti and Girvin, S.M.},
  journal = {PRX Quantum},
  volume = {5},
  issue = {2},
  pages = {020312},
  numpages = {23},
  year = {2024},
  month = {Apr},
  publisher = {American Physical Society},
  doi = {10.1103/PRXQuantum.5.020312},
  url = {https://link.aps.org/doi/10.1103/PRXQuantum.5.020312}
}

@article{qram_giovannetti,
  title = {Quantum Random Access Memory},
  author = {Giovannetti, Vittorio and Lloyd, Seth and Maccone, Lorenzo},
  journal = {Phys. Rev. Lett.},
  volume = {100},
  issue = {16},
  pages = {160501},
  numpages = {4},
  year = {2008},
  month = {Apr},
  publisher = {American Physical Society},
  doi = {10.1103/PhysRevLett.100.160501},
  url = {https://link.aps.org/doi/10.1103/PhysRevLett.100.160501}
}

@article{router_scq_miao,
  title = {Implementation of a Quantum Addressable Router Using Superconducting Qubits},
  author = {Miao, Connie and L\'eger, S\'ebastien and Li, Ziqian and Lee, Gideon and Jiang, Liang and Schuster, David I.},
  journal = {PRX Quantum},
  volume = {6},
  issue = {4},
  pages = {040335},
  numpages = {23},
  year = {2025},
  month = {Nov},
  publisher = {American Physical Society},
  doi = {10.1103/pq3x-cmw9},
  url = {https://link.aps.org/doi/10.1103/pq3x-cmw9}
}

@article{luSchoelkopf2023,
  title = {High-Fidelity Parametric Beamsplitting with a Parity-Protected Converter},
  author = {Lu, Yao and Maiti, Aniket and Garmon, John W. O. and Ganjam, Suhas and Zhang, Yaxing and Claes, Jahan and Frunzio, Luigi and Girvin, Steven M. and Schoelkopf, Robert J.},
  year = {2023},
  month = sep,
  journal = {Nature Communications},
  volume = {14},
  number = {1},
  pages = {5767},
  doi = {10.1038/s41467-023-41104-0},
  urldate = {2024-11-07},
  langid = {english}
}

@article{noise_protected_review_gyenis,
  title = {Moving beyond the Transmon: Noise-Protected Superconducting Quantum Circuits},
  author = {Gyenis, Andr\'as and Di Paolo, Agustin and Koch, Jens and Blais, Alexandre and Houck, Andrew A. and Schuster, David I.},
  journal = {PRX Quantum},
  volume = {2},
  issue = {3},
  pages = {030101},
  numpages = {15},
  year = {2021},
  month = {Sep},
  publisher = {American Physical Society},
  doi = {10.1103/PRXQuantum.2.030101},
  url = {https://link.aps.org/doi/10.1103/PRXQuantum.2.030101}
}

@article{bifluxon_Kalashnikov_2020,
   title={Bifluxon: Fluxon-Parity-Protected Superconducting Qubit},
   volume={1},
   ISSN={2691-3399},
   url={http://dx.doi.org/10.1103/PRXQuantum.1.010307},
   DOI={10.1103/prxquantum.1.010307},
   number={1},
   journal={PRX Quantum},
   publisher={American Physical Society (APS)},
   author={Kalashnikov, Konstantin and Hsieh, Wen Ting and Zhang, Wenyuan and Lu, Wen-Sen and Kamenov, Plamen and Di Paolo, Agustin and Blais, Alexandre and Gershenson, Michael E. and Bell, Matthew},
   year={2020},
   month=sep }

@article{zero_pi,
  title = {Experimental Realization of a Protected Superconducting Circuit Derived from the $0$--$\ensuremath{\pi}$ Qubit},
  author = {Gyenis, Andr\'as and Mundada, Pranav S. and Di Paolo, Agustin and Hazard, Thomas M. and You, Xinyuan and Schuster, David I. and Koch, Jens and Blais, Alexandre and Houck, Andrew A.},
  journal = {PRX Quantum},
  volume = {2},
  issue = {1},
  pages = {010339},
  numpages = {18},
  year = {2021},
  month = {Mar},
  publisher = {American Physical Society},
  doi = {10.1103/PRXQuantum.2.010339},
  url = {https://link.aps.org/doi/10.1103/PRXQuantum.2.010339}
}

@Article{Li2024,
author={Li, Ziqian
and Roy, Tanay
and Rodr{\'i}guez P{\'e}rez, David
and Lee, Kan-Heng
and Kapit, Eliot
and Schuster, David I.},
title={Autonomous error correction of a single logical qubit using two transmons},
journal={Nature Communications},
year={2024},
month={Feb},
day={23},
volume={15},
number={1},
pages={1681},
issn={2041-1723},
doi={10.1038/s41467-024-45858-z},
url={https://doi.org/10.1038/s41467-024-45858-z}
}

@inproceedings{dave_memory_propsal,
author = {Baker, Jonathan M. and Schuster, David I. and Chong, Frederic T.},
title = {Memory-Equipped Quantum Architectures: The Power of Random Access},
year = {2020},
isbn = {9781450380751},
publisher = {Association for Computing Machinery},
address = {New York, NY, USA},
url = {https://doi.org/10.1145/3410463.3414644},
doi = {10.1145/3410463.3414644},
booktitle = {Proc. ACM},
pages = {387–398},
numpages = {12},
location = {Virtual Event, GA, USA},
series = {PACT '20}
}

@article{Manucharyan2009Fluxonium,
  title        = {Fluxonium: Single Cooper-Pair Circuit Free of Charge Offsets},
  author       = {Manucharyan, V. E. and Koch, Jens and Glazman, L. I. and Devoret, M. H.},
  journal      = {Science},
  volume       = {326},
  number       = {5949},
  pages        = {113--116},
  year         = {2009},
  doi          = {10.1126/science.1175552}
}

@ARTICLE{filter_IEEE,
  author={Kirschning, M. and Jansen, R.H.},
  journal={IEEE Transactions on Microwave Theory and Techniques}, 
  title={Accurate Wide-Range Design Equations for the Frequency-Dependent Characteristic of Parallel Coupled Microstrip Lines}, 
  year={1984},
  volume={32},
  number={1},
  pages={83-90},
  doi={10.1109/TMTT.1984.1132616}}

@ARTICLE{balun_IEEE,
  author={Trifunovic, V. and Jokanovic, B.},
  journal={IEEE Transactions on Microwave Theory and Techniques}, 
  title={Review of printed Marchand and double Y baluns: characteristics and application}, 
  year={1994},
  volume={42},
  number={8},
  pages={1454-1462},
  doi={10.1109/22.297806}}

@inproceedings{Venkatesan2003InvestigationOT,
  title={Investigation of the double-y balun for feeding pulsed antennas},
  author={Jaikrishna B. Venkatesan and Waymond R. Scott},
  booktitle={SPIE Defense + Commercial Sensing},
  year={2003},
  url={https://api.semanticscholar.org/CorpusID:53125717}
}

@article{dixit_dark_matter,
  title = {Searching for Dark Matter with a Superconducting Qubit},
  author = {Dixit, Akash V. and Chakram, Srivatsan and He, Kevin and Agrawal, Ankur and Naik, Ravi K. and Schuster, David I. and Chou, Aaron},
  journal = {Phys. Rev. Lett.},
  volume = {126},
  issue = {14},
  pages = {141302},
  numpages = {7},
  year = {2021},
  month = {Apr},
  publisher = {American Physical Society},
  doi = {10.1103/PhysRevLett.126.141302},
  url = {https://link.aps.org/doi/10.1103/PhysRevLett.126.141302}
}

@article{nigg_bbox,
  title = {Black-Box Superconducting Circuit Quantization},
  author = {Nigg, Simon E. and Paik, Hanhee and Vlastakis, Brian and Kirchmair, Gerhard and Shankar, S. and Frunzio, Luigi and Devoret, M. H. and Schoelkopf, R. J. and Girvin, S. M.},
  journal = {Phys. Rev. Lett.},
  volume = {108},
  issue = {24},
  pages = {240502},
  numpages = {5},
  year = {2012},
  month = {Jun},
  publisher = {American Physical Society},
  doi = {10.1103/PhysRevLett.108.240502},
  url = {https://link.aps.org/doi/10.1103/PhysRevLett.108.240502}
}

@misc{oriani2024niobiumcoaxialcavitiesinternal,
      title={Niobium coaxial cavities with internal quality factors exceeding 1.5 billion for circuit quantum electrodynamics}, 
      author={Andrew E. Oriani and Fang Zhao and Tanay Roy and Alexander Anferov and Kevin He and Ankur Agrawal and Riju Banerjee and Srivatsan Chakram and David I. Schuster},
      year={2024},
      eprint={2403.00286},
      archivePrefix={arXiv},
      primaryClass={quant-ph},
      url={https://arxiv.org/abs/2403.00286}, 
}

@article{romanenko_cavities,
  title = {Three-Dimensional Superconducting Resonators at $20$ mK with Photon Lifetimes up to $\ensuremath{\tau}=2$ s},
  author = {Romanenko, A. and Pilipenko, R. and Zorzetti, S. and Frolov, D. and Awida, M. and Belomestnykh, S. and Posen, S. and Grassellino, A.},
  journal = {Phys. Rev. Appl.},
  volume = {13},
  issue = {3},
  pages = {034032},
  numpages = {5},
  year = {2020},
  month = {Mar},
  publisher = {American Physical Society},
  doi = {10.1103/PhysRevApplied.13.034032},
  url = {https://link.aps.org/doi/10.1103/PhysRevApplied.13.034032}
}

@article{milul_cavity_qubit,
  title = {Superconducting Cavity Qubit with Tens of Milliseconds Single-Photon Coherence Time},
  author = {Milul, Ofir and Guttel, Barkay and Goldblatt, Uri and Hazanov, Sergey and Joshi, Lalit M. and Chausovsky, Daniel and Kahn, Nitzan and \ifmmode \mbox{\c{C}}\else \c{C}\fi{}ifty\"urek, Engin and Lafont, Fabien and Rosenblum, Serge},
  journal = {PRX Quantum},
  volume = {4},
  issue = {3},
  pages = {030336},
  numpages = {16},
  year = {2023},
  month = {Sep},
  publisher = {American Physical Society},
  doi = {10.1103/PRXQuantum.4.030336},
  url = {https://link.aps.org/doi/10.1103/PRXQuantum.4.030336}
}

@article{zorin_flux_driven_twpa,
  title = {Flux-Driven Josephson Traveling-Wave Parametric Amplifier},
  author = {Zorin, A.B.},
  journal = {Phys. Rev. Appl.},
  volume = {12},
  issue = {4},
  pages = {044051},
  numpages = {16},
  year = {2019},
  month = {Oct},
  publisher = {American Physical Society},
  doi = {10.1103/PhysRevApplied.12.044051},
  url = {https://link.aps.org/doi/10.1103/PhysRevApplied.12.044051}
}

@Article{Valadares2024,
author={Valadares, Fernando
and Huang, Ni-Ni
and Chu, Kyle Timothy Ng
and Dorogov, Aleksandr
and Chua, Weipin
and Kong, Lingda
and Song, Pengtao
and Gao, Yvonne Y.},
title={On-demand transposition across light-matter interaction regimes in bosonic cQED},
journal={Nature Communications},
year={2024},
month={Jul},
day={10},
volume={15},
number={1},
pages={5816},
issn={2041-1723},
doi={10.1038/s41467-024-50201-7},
url={https://doi.org/10.1038/s41467-024-50201-7}
}

@article{LDPC_review,
  title = {Quantum Low-Density Parity-Check Codes},
  author = {Breuckmann, Nikolas P. and Eberhardt, Jens Niklas},
  journal = {PRX Quantum},
  volume = {2},
  issue = {4},
  pages = {040101},
  numpages = {19},
  year = {2021},
  month = {Oct},
  publisher = {American Physical Society},
  doi = {10.1103/PRXQuantum.2.040101},
  url = {https://link.aps.org/doi/10.1103/PRXQuantum.2.040101}
}

@article{chapman_bs_2023,
  title = {High-On-Off-Ratio Beam-Splitter Interaction for Gates on Bosonically Encoded Qubits},
  author = {Chapman, Benjamin J. and de Graaf, Stijn J. and Xue, Sophia H. and Zhang, Yaxing and Teoh, James and Curtis, Jacob C. and Tsunoda, Takahiro and Eickbusch, Alec and Read, Alexander P. and Koottandavida, Akshay and Mundhada, Shantanu O. and Frunzio, Luigi and Devoret, M.H. and Girvin, S.M. and Schoelkopf, R.J.},
  journal = {PRX Quantum},
  volume = {4},
  issue = {2},
  pages = {020355},
  numpages = {30},
  year = {2023},
  month = {Jun},
  publisher = {American Physical Society},
  doi = {10.1103/PRXQuantum.4.020355},
  url = {https://link.aps.org/doi/10.1103/PRXQuantum.4.020355}
}

@article{rb_knill,
  title = {Randomized benchmarking of quantum gates},
  author = {Knill, E. and Leibfried, D. and Reichle, R. and Britton, J. and Blakestad, R. B. and Jost, J. D. and Langer, C. and Ozeri, R. and Seidelin, S. and Wineland, D. J.},
  journal = {Phys. Rev. A},
  volume = {77},
  issue = {1},
  pages = {012307},
  numpages = {7},
  year = {2008},
  month = {Jan},
  publisher = {American Physical Society},
  doi = {10.1103/PhysRevA.77.012307},
  url = {https://link.aps.org/doi/10.1103/PhysRevA.77.012307}
}

@article{DoubleYbalun_1992,
author = {Trifunovic, V. and Jokanovic, Branka},
year = {1992},
month = {04},
pages = {534 - 535},
title = {Four decade bandwidth uniplanar balun},
volume = {28},
journal = {Electronics Letters},
doi = {10.1049/el:19920337}
}

@Article{Acharya2024,
author={Acharya, Rajeev
and Abanin, Dmitry A.
and Aghababaie-Beni, Laleh
and Aleiner, Igor
and Andersen, Trond I.
and Ansmann, Markus
and Arute, Frank
and Arya, Kunal
and Asfaw, Abraham
and Astrakhantsev, Nikita
and Atalaya, Juan
and Babbush, Ryan
and Bacon, Dave
and Ballard, Brian
and Bardin, Joseph C.
and Bausch, Johannes
and Bengtsson, Andreas
and Bilmes, Alexander
and Blackwell, Sam
and Boixo, Sergio
and Bortoli, Gina
and Bourassa, Alexandre
and Bovaird, Jenna
and Brill, Leon
and Broughton, Michael
and Browne, David A.
and Buchea, Brett
and Buckley, Bob B.
and Buell, David A.
and Burger, Tim
and Burkett, Brian
and Bushnell, Nicholas
and Cabrera, Anthony
and Campero, Juan
and Chang, Hung-Shen
and Chen, Yu
and Chen, Zijun
and Chiaro, Ben
and Chik, Desmond
and Chou, Charina
and Claes, Jahan
and Cleland, Agnetta Y.
and Cogan, Josh
and Collins, Roberto
and Conner, Paul
and Courtney, William
and Crook, Alexander L.
and Curtin, Ben
and Das, Sayan
and Davies, Alex
and De Lorenzo, Laura
and Debroy, Dripto M.
and Demura, Sean
and Devoret, Michel
and Di Paolo, Agustin
and Donohoe, Paul
and Drozdov, Ilya
and Dunsworth, Andrew
and Earle, Clint
and Edlich, Thomas
and Eickbusch, Alec
and Elbag, Aviv Moshe
and Elzouka, Mahmoud
and Erickson, Catherine
and Faoro, Lara
and Farhi, Edward
and Ferreira, Vinicius S.
and Burgos, Leslie Flores
and Forati, Ebrahim
and Fowler, Austin G.
and Foxen, Brooks
and Ganjam, Suhas
and Garcia, Gonzalo
and Gasca, Robert
and Genois, {\'E}lie
and Giang, William
and Gidney, Craig
and Gilboa, Dar
and Gosula, Raja
and Dau, Alejandro Grajales
and Graumann, Dietrich
and Greene, Alex
and Gross, Jonathan A.
and Habegger, Steve
and Hall, John
and Hamilton, Michael C.
and Hansen, Monica
and Harrigan, Matthew P.
and Harrington, Sean D.
and Heras, Francisco J. H.
and Heslin, Stephen
and Heu, Paula
and Higgott, Oscar
and Hill, Gordon
and Hilton, Jeremy
and Holland, George
and Hong, Sabrina
and Huang, Hsin-Yuan
and Huff, Ashley
and Huggins, William J.
and Ioffe, Lev B.
and Isakov, Sergei V.
and Iveland, Justin
and Jeffrey, Evan
and Jiang, Zhang
and Jones, Cody
and Jordan, Stephen
and Joshi, Chaitali
and Juhas, Pavol
and Kafri, Dvir
and Kang, Hui
and Karamlou, Amir H.
and Kechedzhi, Kostyantyn
and Kelly, Julian
and Khaire, Trupti
and Khattar, Tanuj
and Khezri, Mostafa
and Kim, Seon
and Klimov, Paul V.
and Klots, Andrey R.
and Kobrin, Bryce
and Kohli, Pushmeet
and Korotkov, Alexander N.
and Kostritsa, Fedor
and Kothari, Robin
and Kozlovskii, Borislav
and Kreikebaum, John Mark
and Kurilovich, Vladislav D.
and Lacroix, Nathan
and Landhuis, David
and Lange-Dei, Tiano
and Langley, Brandon W.
and Laptev, Pavel
and Lau, Kim-Ming
and Le Guevel, Lo{\"i}ck
and Ledford, Justin
and Lee, Joonho
and Lee, Kenny
and Lensky, Yuri D.
and Leon, Shannon
and Lester, Brian J.
and Li, Wing Yan
and Li, Yin
and Lill, Alexander T.
and Liu, Wayne
and Livingston, William P.
and Locharla, Aditya
and Lucero, Erik
and Lundahl, Daniel
and Lunt, Aaron
and Madhuk, Sid
and Malone, Fionn D.
and Maloney, Ashley
and Mandr{\`a}, Salvatore
and Manyika, James
and Martin, Leigh S.
and Martin, Orion
and Martin, Steven
and Maxfield, Cameron
and McClean, Jarrod R.
and McEwen, Matt
and Meeks, Seneca
and Megrant, Anthony
and Mi, Xiao
and Miao, Kevin C.
and Mieszala, Amanda
and Molavi, Reza
and Molina, Sebastian
and Montazeri, Shirin
and Morvan, Alexis
and Movassagh, Ramis
and Mruczkiewicz, Wojciech
and Naaman, Ofer
and Neeley, Matthew
and Neill, Charles
and Nersisyan, Ani
and Neven, Hartmut
and Newman, Michael
and Ng, Jiun How
and Nguyen, Anthony
and Nguyen, Murray
and Ni, Chia-Hung
and Niu, Murphy Yuezhen
and O'Brien, Thomas E.
and Oliver, William D.
and Opremcak, Alex
and Ottosson, Kristoffer
and Petukhov, Andre
and Pizzuto, Alex
and Platt, John
and Potter, Rebecca
and Pritchard, Orion
and Pryadko, Leonid P.
and Quintana, Chris
and Ramachandran, Ganesh
and Reagor, Matthew J.
and Redding, John
and Rhodes, David M.
and Roberts, Gabrielle
and Rosenberg, Eliott
and Rosenfeld, Emma
and Roushan, Pedram
and Rubin, Nicholas C.
and Saei, Negar
and Sank, Daniel
and Sankaragomathi, Kannan
and Satzinger, Kevin J.
and Schurkus, Henry F.
and Schuster, Christopher
and Senior, Andrew W.
and Shearn, Michael J.
and Shorter, Aaron
and Shutty, Noah
and Shvarts, Vladimir
and Singh, Shraddha
and Sivak, Volodymyr
and Skruzny, Jindra
and Small, Spencer
and Smelyanskiy, Vadim
and Smith, W. Clarke
and Somma, Rolando D.
and Springer, Sofia
and Sterling, George
and Strain, Doug
and Suchard, Jordan
and Szasz, Aaron
and Sztein, Alex
and Thor, Douglas
and Torres, Alfredo
and Torunbalci, M. Mert
and Vaishnav, Abeer
and Vargas, Justin
and Vdovichev, Sergey
and Vidal, Guifre
and Villalonga, Benjamin
and Heidweiller, Catherine Vollgraff
and Waltman, Steven
and Wang, Shannon X.
and Ware, Brayden
and Weber, Kate
and Weidel, Travis
and White, Theodore
and Wong, Kristi
and Woo, Bryan W. K.
and Xing, Cheng
and Yao, Z. Jamie
and Yeh, Ping
and Ying, Bicheng
and Yoo, Juhwan
and Yosri, Noureldin
and Young, Grayson
and Zalcman, Adam
and Zhang, Yaxing
and Zhu, Ningfeng
and Zobrist, Nicholas
and AI, Google Quantum
and {Collaborators}},
title={Quantum error correction below the surface code threshold},
journal={Nature},
year={2025},
month={Feb},
day={01},
volume={638},
number={8052},
pages={920-926},
issn={1476-4687},
doi={10.1038/s41586-024-08449-y},
url={https://doi.org/10.1038/s41586-024-08449-y}
}

@misc{huang2025fastsidebandcontrolweakly,
      title={Fast Sideband Control of a Weakly Coupled Multimode Bosonic Memory}, 
      author={Jordan Huang and Thomas J. DiNapoli and Gavin Rockwood and Ming Yuan and Prathyankara Narasimhan and Eesh Gupta and Mustafa Bal and Francesco Crisa and Sabrina Garattoni and Yao Lu and Liang Jiang and Srivatsan Chakram},
      year={2025},
      eprint={2503.10623},
      archivePrefix={arXiv},
      primaryClass={quant-ph},
      url={https://arxiv.org/abs/2503.10623}, 
}

@article{heeres_grape_2017,
   title={Implementing a universal gate set on a logical qubit encoded in an oscillator},
   volume={8},
   ISSN={2041-1723},
   url={http://dx.doi.org/10.1038/s41467-017-00045-1},
   DOI={10.1038/s41467-017-00045-1},
   number={1},
   journal={Nature Communications},
   publisher={Springer Science and Business Media LLC},
   author={Heeres, Reinier W. and Reinhold, Philip and Ofek, Nissim and Frunzio, Luigi and Jiang, Liang and Devoret, Michel H. and Schoelkopf, Robert J.},
   year={2017},
   month=jul }

@article{pechal_sideband_transmon_cavity,
  title = {Microwave-Controlled Generation of Shaped Single Photons in Circuit Quantum Electrodynamics},
  author = {Pechal, M. and Huthmacher, L. and Eichler, C. and Zeytino\ifmmode \breve{g}\else \u{g}\fi{}lu, S. and Abdumalikov, A. A. and Berger, S. and Wallraff, A. and Filipp, S.},
  journal = {Phys. Rev. X},
  volume = {4},
  issue = {4},
  pages = {041010},
  numpages = {9},
  year = {2014},
  month = {Oct},
  publisher = {American Physical Society},
  doi = {10.1103/PhysRevX.4.041010},
  url = {https://link.aps.org/doi/10.1103/PhysRevX.4.041010}
}

@article{PhysRevLett.127.107701_vatsan_highQ,
  title = {Seamless High-$Q$ Microwave Cavities for Multimode Circuit Quantum Electrodynamics},
  author = {Chakram, Srivatsan and Oriani, Andrew E. and Naik, Ravi K. and Dixit, Akash V. and He, Kevin and Agrawal, Ankur and Kwon, Hyeokshin and Schuster, David I.},
  journal = {Phys. Rev. Lett.},
  volume = {127},
  issue = {10},
  pages = {107701},
  numpages = {6},
  year = {2021},
  month = {Aug},
  publisher = {American Physical Society},
  doi = {10.1103/PhysRevLett.127.107701},
  url = {https://link.aps.org/doi/10.1103/PhysRevLett.127.107701}
}

@Article{Chakram2022,
author={Chakram, Srivatsan
and He, Kevin
and Dixit, Akash V.
and Oriani, Andrew E.
and Naik, Ravi K.
and Leung, Nelson
and Kwon, Hyeokshin
and Ma, Wen-Long
and Jiang, Liang
and Schuster, David I.},
title={Multimode photon blockade},
journal={Nature Physics},
year={2022},
month={Aug},
day={01},
volume={18},
number={8},
pages={879-884},
issn={1745-2481},
doi={10.1038/s41567-022-01630-y},
url={https://doi.org/10.1038/s41567-022-01630-y}
}

@Article{Eickbusch2022,
author={Eickbusch, Alec
and Sivak, Volodymyr
and Ding, Andy Z.
and Elder, Salvatore S.
and Jha, Shantanu R.
and Venkatraman, Jayameenakshi
and Royer, Baptiste
and Girvin, S. M.
and Schoelkopf, Robert J.
and Devoret, Michel H.},
title={Fast universal control of an oscillator with weak dispersive coupling to a qubit},
journal={Nature Physics},
year={2022},
month={Dec},
day={01},
volume={18},
number={12},
pages={1464-1469},
issn={1745-2481},
doi={10.1038/s41567-022-01776-9},
url={https://doi.org/10.1038/s41567-022-01776-9}
}

@article{girvin_strategies_tradeoffs_memory_2024,
  title = {Strategies and Trade-Offs for Controllability and Memory Time of Ultra-High-Quality Microwave Cavities in Circuit Quantum Electrodynamics},
  author = {Pietik\"ainen, Iivari and \ifmmode \check{C}\else \v{C}\fi{}ernot\'{\i}k, Ond\ifmmode \check{r}\else \v{r}\fi{}ej and Eickbusch, Alec and Maiti, Aniket and Garmon, John W.O. and Filip, Radim and Girvin, Steven M.},
  journal = {PRX Quantum},
  volume = {5},
  issue = {4},
  pages = {040307},
  numpages = {30},
  year = {2024},
  month = {Oct},
  publisher = {American Physical Society},
  doi = {10.1103/PRXQuantum.5.040307},
  url = {https://link.aps.org/doi/10.1103/PRXQuantum.5.040307}
}

@article{bravyi_decoding_surface_code_2014,
  title = {Efficient algorithms for maximum likelihood decoding in the surface code},
  author = {Bravyi, Sergey and Suchara, Martin and Vargo, Alexander},
  journal = {Phys. Rev. A},
  volume = {90},
  issue = {3},
  pages = {032326},
  numpages = {15},
  year = {2014},
  month = {Sep},
  publisher = {American Physical Society},
  doi = {10.1103/PhysRevA.90.032326},
  url = {https://link.aps.org/doi/10.1103/PhysRevA.90.032326}
}

@article{Criger2018multipathsummation,
  doi = {10.22331/q-2018-10-19-102},
  url = {https://doi.org/10.22331/q-2018-10-19-102},
  title = {Multi-path {S}ummation for {D}ecoding 2{D} {T}opological {C}odes},
  author = {Criger, Ben and Ashraf, Imran},
  journal = {{Quantum}},
  issn = {2521-327X},
  publisher = {{Verein zur F{\"{o}}rderung des Open Access Publizierens in den Quantenwissenschaften}},
  volume = {2},
  pages = {102},
  month = oct,
  year = {2018}
}

@Article{Kuo2022,
author={Kuo, Kao-Yueh
and Lai, Ching-Yi},
title={Exploiting degeneracy in belief propagation decoding of quantum codes},
journal={npj Quantum Information},
year={2022},
month={Sep},
day={14},
volume={8},
number={1},
pages={111},
issn={2056-6387},
doi={10.1038/s41534-022-00623-2},
url={https://doi.org/10.1038/s41534-022-00623-2}
}

@InProceedings{Duckering2020,
 author={Duckering, Casey and Baker, Jonathan M. and Schuster, David I. and Chong, Frederic T.},
  booktitle={{IEEE/ACM Int. Symp. on Microarchitecture (MICRO)}}, 
  title={Virtualized Logical Qubits: A 2.5D Architecture for Error-Corrected Quantum Computing}, 
  year={2020},
  volume={},
  number={},
  pages={173-185},
  doi={10.1109/MICRO50266.2020.00026}}

@article{wootton_2012_high_threshold,
  title = {High Threshold Error Correction for the Surface Code},
  author = {Wootton, James R. and Loss, Daniel},
  journal = {Phys. Rev. Lett.},
  volume = {109},
  issue = {16},
  pages = {160503},
  numpages = {5},
  year = {2012},
  month = {Oct},
  publisher = {American Physical Society},
  doi = {10.1103/PhysRevLett.109.160503},
  url = {https://link.aps.org/doi/10.1103/PhysRevLett.109.160503}
}

@unpublished{Lu2025,
  author = {Y. Lu and others},
  title = {Systematic construction of time-dependent {Hamiltonians} for microwave-driven {Josephson} circuits},
  note = {in preparation},
}

@misc{maiti2025linearquantumcouplerclean,
      title={A Linear Quantum Coupler for Clean Bosonic Control}, 
      author={Aniket Maiti and John W. O. Garmon and Yao Lu and Alessandro Miano and Luigi Frunzio and Robert J. Schoelkopf},
      year={2025},
      eprint={2501.18025},
      archivePrefix={arXiv},
      primaryClass={quant-ph},
      url={https://arxiv.org/abs/2501.18025}, 
}

@Article{Campagne-Ibarcq2020,
author={Campagne-Ibarcq, P.
and Eickbusch, A.
and Touzard, S.
and Zalys-Geller, E.
and Frattini, N. E.
and Sivak, V. V.
and Reinhold, P.
and Puri, S.
and Shankar, S.
and Schoelkopf, R. J.
and Frunzio, L.
and Mirrahimi, M.
and Devoret, M. H.},
title={Quantum error correction of a qubit encoded in grid states of an oscillator},
journal={Nature},
year={2020},
month={Aug},
day={01},
volume={584},
number={7821},
pages={368-372},
issn={1476-4687},
doi={10.1038/s41586-020-2603-3},
url={https://doi.org/10.1038/s41586-020-2603-3}
}

@Article{Sivak2023,
author={Sivak, V. V.
and Eickbusch, A.
and Royer, B.
and Singh, S.
and Tsioutsios, I.
and Ganjam, S.
and Miano, A.
and Brock, B. L.
and Ding, A. Z.
and Frunzio, L.
and Girvin, S. M.
and Schoelkopf, R. J.
and Devoret, M. H.},
title={Real-time quantum error correction beyond break-even},
journal={Nature},
year={2023},
month={Apr},
day={01},
volume={616},
number={7955},
pages={50-55},
issn={1476-4687},
doi={10.1038/s41586-023-05782-6},
url={https://doi.org/10.1038/s41586-023-05782-6}
}

@Article{Bluvstein2024,
author={Bluvstein, Dolev
and Evered, Simon J.
and Geim, Alexandra A.
and Li, Sophie H.
and Zhou, Hengyun
and Manovitz, Tom
and Ebadi, Sepehr
and Cain, Madelyn
and Kalinowski, Marcin
and Hangleiter, Dominik
and Bonilla Ataides, J. Pablo
and Maskara, Nishad
and Cong, Iris
and Gao, Xun
and Sales Rodriguez, Pedro
and Karolyshyn, Thomas
and Semeghini, Giulia
and Gullans, Michael J.
and Greiner, Markus
and Vuleti{\'{c}}, Vladan
and Lukin, Mikhail D.},
title={Logical quantum processor based on reconfigurable atom arrays},
journal={Nature},
year={2024},
month={Feb},
day={01},
volume={626},
number={7997},
pages={58-65},
issn={1476-4687},
doi={10.1038/s41586-023-06927-3},
url={https://doi.org/10.1038/s41586-023-06927-3}
}

@article{ryan_anderson_ftqec_2021,
  title = {Realization of Real-Time Fault-Tolerant Quantum Error Correction},
  author = {Ryan-Anderson, C. and Bohnet, J. G. and Lee, K. and Gresh, D. and Hankin, A. and Gaebler, J. P. and Francois, D. and Chernoguzov, A. and Lucchetti, D. and Brown, N. C. and Gatterman, T. M. and Halit, S. K. and Gilmore, K. and Gerber, J. A. and Neyenhuis, B. and Hayes, D. and Stutz, R. P.},
  journal = {Phys. Rev. X},
  volume = {11},
  issue = {4},
  pages = {041058},
  numpages = {29},
  year = {2021},
  month = {Dec},
  publisher = {American Physical Society},
  doi = {10.1103/PhysRevX.11.041058},
  url = {https://link.aps.org/doi/10.1103/PhysRevX.11.041058}
}

@Article{Takeda2022,
author={Takeda, Kenta
and Noiri, Akito
and Nakajima, Takashi
and Kobayashi, Takashi
and Tarucha, Seigo},
title={Quantum error correction with silicon spin qubits},
journal={Nature},
year={2022},
month={Aug},
day={01},
volume={608},
number={7924},
pages={682-686},
issn={1476-4687},
doi={10.1038/s41586-022-04986-6},
url={https://doi.org/10.1038/s41586-022-04986-6}
}

@Article{Ni2023,
author={Ni, Zhongchu
and Li, Sai
and Deng, Xiaowei
and Cai, Yanyan
and Zhang, Libo
and Wang, Weiting
and Yang, Zhen-Biao
and Yu, Haifeng
and Yan, Fei
and Liu, Song
and Zou, Chang-Ling
and Sun, Luyan
and Zheng, Shi-Biao
and Xu, Yuan
and Yu, Dapeng},
title={Beating the break-even point with a discrete-variable-encoded logical qubit},
journal={Nature},
year={2023},
month={Apr},
day={01},
volume={616},
number={7955},
pages={56-60},
issn={1476-4687},
doi={10.1038/s41586-023-05784-4},
url={https://doi.org/10.1038/s41586-023-05784-4}
}

@article{Herr_2017,
doi = {10.1088/1367-2630/aa5709},
url = {https://dx.doi.org/10.1088/1367-2630/aa5709},
year = {2017},
month = {jan},
publisher = {IOP Publishing},
volume = {19},
number = {1},
pages = {013034},
author = {Herr, Daniel and Nori, Franco and Devitt, Simon J},
title = {Lattice surgery translation for quantum computation},
journal = {New Journal of Physics}
}

@article{Horsman_2012,
doi = {10.1088/1367-2630/14/12/123011},
url = {https://doi.org/10.1088/1367-2630/14/12/123011},
year = {2012},
month = {dec},
publisher = {IOP Publishing},
volume = {14},
number = {12},
pages = {123011},
author = {Horsman, Dominic and Fowler, Austin G and Devitt, Simon and Meter, Rodney Van},
title = {Surface code quantum computing by lattice surgery},
journal = {New Journal of Physics}
}

@article{erasure_koottandavida_2024,
  title = {Erasure Detection of a Dual-Rail Qubit Encoded in a Double-Post Superconducting Cavity},
  author = {Koottandavida, Akshay and Tsioutsios, Ioannis and Kargioti, Aikaterini and Smith, Cassady R. and Joshi, Vidul R. and Dai, Wei and Teoh, James D. and Curtis, Jacob C. and Frunzio, Luigi and Schoelkopf, Robert J. and Devoret, Michel H.},
  journal = {Phys. Rev. Lett.},
  volume = {132},
  issue = {18},
  pages = {180601},
  numpages = {7},
  year = {2024},
  month = {May},
  publisher = {American Physical Society},
  doi = {10.1103/PhysRevLett.132.180601},
  url = {https://link.aps.org/doi/10.1103/PhysRevLett.132.180601}
}

@Article{Chou2024_dual_rail,
author={Chou, Kevin S.
and Shemma, Tali
and McCarrick, Heather
and Chien, Tzu-Chiao
and Teoh, James D.
and Winkel, Patrick
and Anderson, Amos
and Chen, Jonathan
and Curtis, Jacob C.
and de Graaf, Stijn J.
and Garmon, John W. O.
and Gudlewski, Benjamin
and Kalfus, William D.
and Keen, Trevor
and Khedkar, Nishaad
and Lei, Chan U.
and Liu, Gangqiang
and Lu, Pinlei
and Lu, Yao
and Maiti, Aniket
and Mastalli-Kelly, Luke
and Mehta, Nitish
and Mundhada, Shantanu O.
and Narla, Anirudh
and Noh, Taewan
and Tsunoda, Takahiro
and Xue, Sophia H.
and Yuan, Joseph O.
and Frunzio, Luigi
and Aumentado, Jos{\'e}
and Puri, Shruti
and Girvin, Steven M.
and Moseley, S. Harvey
and Schoelkopf, Robert J.},
title={A superconducting dual-rail cavity qubit with erasure-detected logical measurements},
journal={Nature Physics},
year={2024},
month={Sep},
day={01},
volume={20},
number={9},
pages={1454-1460},
issn={1745-2481},
doi={10.1038/s41567-024-02539-4},
url={https://doi.org/10.1038/s41567-024-02539-4}
}

@Article{Ofek2016,
author={Ofek, Nissim
and Petrenko, Andrei
and Heeres, Reinier
and Reinhold, Philip
and Leghtas, Zaki
and Vlastakis, Brian
and Liu, Yehan
and Frunzio, Luigi
and Girvin, S. M.
and Jiang, L.
and Mirrahimi, Mazyar
and Devoret, M. H.
and Schoelkopf, R. J.},
title={Extending the lifetime of a quantum bit with error correction in superconducting circuits},
journal={Nature},
year={2016},
month={Aug},
day={01},
volume={536},
number={7617},
pages={441-445},
issn={1476-4687},
doi={10.1038/nature18949},
url={https://doi.org/10.1038/nature18949}
}

@article{
willke_2018_hyperfine,
author = {Philip Willke  and Yujeong Bae  and Kai Yang  and Jose L. Lado  and Alejandro Ferrón  and Taeyoung Choi  and Arzhang Ardavan  and Joaquín Fernández-Rossier  and Andreas J. Heinrich  and Christopher P. Lutz },
title = {Hyperfine interaction of individual atoms on a surface},
journal = {Science},
volume = {362},
number = {6412},
pages = {336-339},
year = {2018},
doi = {10.1126/science.aat7047},
URL = {https://www.science.org/doi/abs/10.1126/science.aat7047}}

@article{SNAP_2015,
  title = {Universal control of an oscillator with dispersive coupling to a qubit},
  author = {Krastanov, Stefan and Albert, Victor V. and Shen, Chao and Zou, Chang-Ling and Heeres, Reinier W. and Vlastakis, Brian and Schoelkopf, Robert J. and Jiang, Liang},
  journal = {Phys. Rev. A},
  volume = {92},
  issue = {4},
  pages = {040303},
  numpages = {5},
  year = {2015},
  month = {Oct},
  publisher = {American Physical Society},
  doi = {10.1103/PhysRevA.92.040303},
  url = {https://link.aps.org/doi/10.1103/PhysRevA.92.040303}
}

@Article{Reiner2024,
author={Reiner, J.
and Chung, Y.
and Misha, S. H.
and Lehner, C.
and Moehle, C.
and Poulos, D.
and Monir, S.
and Charde, K. J.
and Macha, P.
and Kranz, L.
and Thorvaldson, I.
and Thorgrimsson, B.
and Keith, D.
and Hsueh, Y. L.
and Rahman, R.
and Gorman, S. K.
and Keizer, J. G.
and Simmons, M. Y.},
title={High-fidelity initialization and control of electron and nuclear spins in a four-qubit register},
journal={Nature Nanotechnology},
year={2024},
month={May},
day={01},
volume={19},
number={5},
pages={605-611},
issn={1748-3395},
doi={10.1038/s41565-023-01596-9},
url={https://doi.org/10.1038/s41565-023-01596-9}
}

@Article{Chou2018,
author={Chou, Kevin S.
and Blumoff, Jacob Z.
and Wang, Christopher S.
and Reinhold, Philip C.
and Axline, Christopher J.
and Gao, Yvonne Y.
and Frunzio, L.
and Devoret, M. H.
and Jiang, Liang
and Schoelkopf, R. J.},
title={Deterministic teleportation of a quantum gate between two logical qubits},
journal={Nature},
year={2018},
month={Sep},
day={01},
volume={561},
number={7723},
pages={368-373},
issn={1476-4687},
doi={10.1038/s41586-018-0470-y},
url={https://doi.org/10.1038/s41586-018-0470-y}
}

@Article{Cupertino2024,
author={Cupertino, Andrea
and Shin, Dongil
and Guo, Leo
and Steeneken, Peter G.
and Bessa, Miguel A.
and Norte, Richard A.},
title={Centimeter-scale nanomechanical resonators with low dissipation},
journal={Nature Communications},
year={2024},
month={May},
day={18},
volume={15},
number={1},
pages={4255},
issn={2041-1723},
doi={10.1038/s41467-024-48183-7},
url={https://doi.org/10.1038/s41467-024-48183-7}
}

@article{marios_bosonic_QEC_2016,
  title = {New Class of Quantum Error-Correcting Codes for a Bosonic Mode},
  author = {Michael, Marios H. and Silveri, Matti and Brierley, R. T. and Albert, Victor V. and Salmilehto, Juha and Jiang, Liang and Girvin, S. M.},
  journal = {Phys. Rev. X},
  volume = {6},
  issue = {3},
  pages = {031006},
  numpages = {26},
  year = {2016},
  month = {Jul},
  publisher = {American Physical Society},
  doi = {10.1103/PhysRevX.6.031006},
  url = {https://link.aps.org/doi/10.1103/PhysRevX.6.031006}
}

@Article{Li2024_autonomous,
author={Li, Ziqian
and Roy, Tanay
and Lu, Yao
and Kapit, Eliot
and Schuster, David I.},
title={Autonomous stabilization with programmable stabilized state},
journal={Nature Communications},
year={2024},
month={Aug},
day={14},
volume={15},
number={1},
pages={6978},
issn={2041-1723},
doi={10.1038/s41467-024-51262-4},
url={https://doi.org/10.1038/s41467-024-51262-4}
}

@article{
teoh_2023_dual_rail,
author = {James D. Teoh  and Patrick Winkel  and Harshvardhan K. Babla  and Benjamin J. Chapman  and Jahan Claes  and Stijn J. de Graaf  and John W. O. Garmon  and William D. Kalfus  and Yao Lu  and Aniket Maiti  and Kaavya Sahay  and Neel Thakur  and Takahiro Tsunoda  and Sophia H. Xue  and Luigi Frunzio  and Steven M. Girvin  and Shruti Puri  and Robert J. Schoelkopf },
title = {Dual-rail encoding with superconducting cavities},
journal = {Proceedings of the National Academy of Sciences},
volume = {120},
number = {41},
pages = {e2221736120},
year = {2023},
doi = {10.1073/pnas.2221736120},
URL = {https://www.pnas.org/doi/abs/10.1073/pnas.2221736120}}

@article{Daul_rail_levine_2024,
  title = {Demonstrating a Long-Coherence Dual-Rail Erasure Qubit Using Tunable Transmons},
  author = {Levine, H. and Haim, A. and Hung, J. S. C. and Alidoust, N. and Kalaee, M. and DeLorenzo, L. and Wollack, E. A. and Arrangoiz-Arriola, P. and Khalajhedayati, A. and Sanil, R. and Moradinejad, H. and Vaknin, Y. and Kubica, A. and Hover, D. and Aghaeimeibodi, S. and Alcid, J. A. and Baek, C. and Barnett, J. and Bawdekar, K. and Bienias, P. and Carson, H. A. and Chen, C. and Chen, L. and Chinkezian, H. and Chisholm, E. M. and Clifford, A. and Cosmic, R. and Crisosto, N. and Dalzell, A. M. and Davis, E. and D'Ewart, J. M. and Diez, S. and D'Souza, N. and Dumitrescu, P. T. and Elkhouly, E. and Fang, M. T. and Fang, Y. and Flammia, S. and Fling, M. J. and Garcia, G. and Gharzai, M. K. and Gorshkov, A. V. and Gray, M. J. and Grimberg, S. and Grimsmo, A. L. and Hann, C. T. and He, Y. and Heidel, S. and Howell, S. and Hunt, M. and Iverson, J. and Jarrige, I. and Jiang, L. and Jones, W. M. and Karabalin, R. and Karalekas, P. J. and Keller, A. J. and Lasi, D. and Lee, M. and Ly, V. and MacCabe, G. and Mahuli, N. and Marcaud, G. and Matheny, M. H. and McArdle, S. and McCabe, G. and Merton, G. and Miles, C. and Milsted, A. and Mishra, A. and Moncelsi, L. and Naghiloo, M. and Noh, K. and Oblepias, E. and Ortuno, G. and Owens, J. C. and Pagdilao, J. and Panduro, A. and Paquette, J.-P. and Patel, R. N. and Peairs, G. and Perello, D. J. and Peterson, E. C. and Ponte, S. and Putterman, H. and Refael, G. and Reinhold, P. and Resnick, R. and Reyna, O. A. and Rodriguez, R. and Rose, J. and Rubin, A. H. and Runyan, M. and Ryan, C. A. and Sahmoud, A. and Scaffidi, T. and Shah, B. and Siavoshi, S. and Sivarajah, P. and Skogland, T. and Su, C.-J. and Swenson, L. J. and Sylvia, J. and Teo, S. M. and Tomada, A. and Torlai, G. and Wistrom, M. and Zhang, K. and Zuk, I. and Clerk, A. A. and Brand\~ao, F. G. S. L. and Retzker, A. and Painter, O.},
  journal = {Phys. Rev. X},
  volume = {14},
  issue = {1},
  pages = {011051},
  numpages = {21},
  year = {2024},
  month = {Mar},
  publisher = {American Physical Society},
  doi = {10.1103/PhysRevX.14.011051},
  url = {https://link.aps.org/doi/10.1103/PhysRevX.14.011051}
}

@article{koch2007transmon,
  title = {Charge-insensitive qubit design derived from the {Cooper} pair box},
  author = {Koch, Jens and Yu, Terri M. and Gambetta, Jay and Houck, A. A. and Schuster, D. I. and Majer, J. and Blais, Alexandre and Devoret, M. H. and Girvin, S. M. and Schoelkopf, R. J.},
  journal = {Physical Review A},
  volume = {76},
  issue = {4},
  pages = {042319},
  numpages = {19},
  year = {2007},
  month = {Oct},
  publisher = {American Physical Society},
  doi = {10.1103/PhysRevA.76.042319}
}

@article{Johansson_2012_qutip,
   title={QuTiP: An open-source Python framework for the dynamics of open quantum systems},
   volume={183},
   ISSN={0010-4655},
   url={http://dx.doi.org/10.1016/j.cpc.2012.02.021},
   DOI={10.1016/j.cpc.2012.02.021},
   number={8},
   journal={Computer Physics Communications},
   publisher={Elsevier BV},
   author={Johansson, J.R. and Nation, P.D. and Nori, Franco},
   year={2012},
   month=aug, pages={1760–1772} }

@article{Stefanazzi_2022_qick,
   title={The QICK (Quantum Instrumentation Control Kit): Readout and control for qubits and detectors},
   volume={93},
   ISSN={1089-7623},
   url={http://dx.doi.org/10.1063/5.0076249},
   DOI={10.1063/5.0076249},
   number={4},
   journal={Review of Scientific Instruments},
   publisher={AIP Publishing},
   author={Stefanazzi, Leandro and Treptow, Kenneth and Wilcer, Neal and Stoughton, Chris and Bradford, Collin and Uemura, Sho and Zorzetti, Silvia and Montella, Salvatore and Cancelo, Gustavo and Sussman, Sara and Houck, Andrew and Saxena, Shefali and Arnaldi, Horacio and Agrawal, Ankur and Zhang, Helin and Ding, Chunyang and Schuster, David I.},
   year={2022},
   month=apr }

@misc{zhao2025fluxtunablecavitydarkmatter,
      title={A Flux-Tunable cavity for Dark matter detection}, 
      author={Fang Zhao and Ziqian Li and Akash V. Dixit and Tanay Roy and Andrei Vrajitoarea and Riju Banerjee and Alexander Anferov and Kan-Heng Lee and David I. Schuster and Aaron Chou},
      year={2025},
      eprint={2501.06882},
      archivePrefix={arXiv},
      primaryClass={quant-ph},
      url={https://arxiv.org/abs/2501.06882}, 
}

@article{PhysRevResearch.2.033447,
  title = {Benchmarking the noise sensitivity of different parametric two-qubit gates in a single superconducting quantum computing platform},
  author = {Ganzhorn, M. and Salis, G. and Egger, D. J. and Fuhrer, A. and Mergenthaler, M. and M\"uller, C. and M\"uller, P. and Paredes, S. and Pechal, M. and Werninghaus, M. and Filipp, S.},
  journal = {Phys. Rev. Res.},
  volume = {2},
  issue = {3},
  pages = {033447},
  numpages = {18},
  year = {2020},
  month = {Sep},
  publisher = {American Physical Society},
  doi = {10.1103/PhysRevResearch.2.033447},
  url = {https://link.aps.org/doi/10.1103/PhysRevResearch.2.033447}
}
\end{document}